\documentclass[pdflatex,sn-mathphys-num]{sn-jnl}

\usepackage{graphicx}%
\usepackage{multirow}%
\usepackage{amsmath,amssymb,amsfonts}%
\usepackage{amsthm}%
\usepackage{mathrsfs}%
\usepackage[title]{appendix}%
\usepackage{xcolor}%
\usepackage{textcomp}%
\usepackage{manyfoot}%
\usepackage{booktabs}%
\usepackage{algorithm}%
\usepackage{algorithmicx}%
\usepackage{algpseudocode}%
\usepackage{listings}%
\usepackage{soul}
\definecolor{newcolor}{rgb}{.8,.349,.1}

\newcommand\vp[1]{{\color{red}\textbf{#1}}}

\theoremstyle{thmstyleone}%
\theoremstyle{thmstyletwo}%

\theoremstyle{thmstylethree}%

\begin{document}

\title[Article Title]{Planetary Radio Interferometry and Doppler Experiment as an operational component of the Jupiter Icy Moons Explorer mission}


\author*[1]{\fnm{Vidhya} \sur{Pallichadath}}\email{v.pallichadath@tudelft.nl, ORCID: 0000-0003-0678-4814}

\author[1]{\fnm{Dominic} \sur{Dirkx}}\email{(d.dirkx@tudelft.nl, ORCID: 0000-0003-2069-0603)}

\author[1]{\fnm{Marie~S.} \sur{Fayolle}}\email{(m.s.fayolle-chambe@tudelft.nl, ORCID: 0000-0003-4407-5031)}

\author[2]{\fnm{S\'{a}ndor} \sur{Frey}}\email{(frey.sandor@csfk.org, ORCID: 0000-0003-3079-1889}

\author[1,3]{\fnm{Leonid} \sur{I. Gurvits}}\email{lgurvits@jive.eu, ORCID: 0000-0002-0694-2459}

\author[3,4]{\fnm{Paul} \sur{Boven}}\email{boven@jive.eu, ORCID: }

\author[3]{\fnm{Giuseppe} \sur{Cim\`{o}}}
\email{(cimo@jive.eu, ORCID: 0000-0002-1167-7565)}

\author[2]{\fnm{Judit} \sur{Fogasy}}\email{(fogasy.judit@csfk.org, ORCID: 0000-0003-0003-000X)}

\author[5]{\fnm{Guifr\'{e}} \sur{Molera~Calv\'{e}s}}\email{(guifre.moleracalves@utas.edu.au, ORCID: 0000-0001-8819-0651)}

\author[2]{\fnm{Krisztina} \sur{Perger}}\email{(perger.krisztina@csfk.org, ORCID: 0000-0002-6044-6069)}

\author[3]{\fnm{N.~Masdiana} \sur{Md~Said}}\email{(said@jive.eu, ORCID: 0000-0002-7147-7039)}

\author[1]{\fnm{Bert~L.A.} \sur{Vermeersen}}\email{(l.l.a.vermeersen@tudelft.nl, ORCID: 0000-0002-6329-5972)}

\affil*[1]{\orgdiv{Faculty of Aerospace Engineering}, \orgname{Delft University of Technology}, \orgaddress{\street{Kluyverweg 1}, \city{Delft}, \postcode{2629 HS}, \country{The Netherlands}}}

\affil[2]{\orgdiv{Konkoly Observatory}, \orgname{ELKH Research Centre for Astronomy and Earth Sciences, MTA Centre of Excellence}, \orgaddress{\street{Konkoly Thege M. \'{u}t 15-17}, \city{Budapest}, \postcode{H--1121}, \country{Hungary}}}

\affil[3]{\orgname{Joint Institute for VLBI ERIC}, \orgaddress{\street{Oude Hoogeveensedijk~4}, \city{Dwingeloo}, \postcode{7991~PD}, \country{The Netherlands}}}

\affil[4]{\orgdiv{Leiden Observatory}, \orgname{Leiden University}, \orgaddress{\street{Box 9513}, \city{Leiden}, \postcode{2300 RA}, \country{The Netherlands}}}

\affil[5]{\orgdiv{Physics discipline, School of Natural Sciences}, \orgname{University of Tasmania}, \orgaddress{\street{Street}, \city{Hobart}, \postcode{TAS~7000}, \state{Tasmania}, \country{Australia}}}


\abstract{\ 

We present an overview of the operations and engineering interface for Planetary Radio Interferometry and Doppler Experiment (PRIDE) radio astronomy observations as a scientific component of the ESA's Jupiter Icy Moons Explorer (JUICE) mission, as well as other prospective planetary and space science missions. The article discusses advanced scheduling and planning methods that make it possible to create observing schedules for observations of specific spacecraft in concurrence with observations of natural radio sources. In order to put this into practice and find suitable natural background calibrator sources for PRIDE of JUICE mission, we developed planning and scheduling software. The conventional scheduling software 
for natural celestial radio sources 
is not set up to include spacecraft as observation targets in the necessary control files. Therefore, difficulties already arise during observation planning. We report on the development of new and the adaptation of existing routines used in astrophysical and geodetic VLBI for satellite scheduling and planning. The analysis of the PRIDE science observations led to improved observational planning, and the mission's scheduling methodologies were studied using a systems engineering approach. In addition, we highlighted the new procedures, like finding charts for selecting calibrator radio sources over a range of frequency bands 
and the outcomes of those strategies for science operation planning. A simulation of the flyby of Venus during the cruise phase of the JUICE spacecraft, based on the Tudat software, is also presented, resulting in a promising opportunity to test PRIDE techniques and evaluate the improvements that PRIDE observables can make to natural bodies' ephemerides. The first K$_{a}$-band (32~GHz) observations of the ESA's BepiColombo by a radio telescope in the VLBI network, which employs a similar radio communications system as JUICE, were also demonstrated as a test case. The primary objective of these activities is to serve as a practice run for the 
upcoming operational PRIDE JUICE operations. We showcase the capabilities of the planning and scheduling software for 
other space missions.}




\maketitle

\section{Introduction}\label{intro}

\ \\
Very Long Baseline Interferometry (VLBI) was developed as a radio astronomical technique in the 1960s to sharpen angular resolution for studies of natural celestial radio sources primarily in the interests of astrophysical studies \citep[][Chapter~1]{TMS-2017}. The method turned out to be so powerful and versatile that its applications had been developed quickly in other areas of astronomy and space science, ranging from astrometry and geodesy \citep[][Chapter~12]{TMS-2017} to fundamental physics \citep[e.g.,][]{shapiro2004measurement,Fomlaont+2009ApJ}, studies in heliophysics and interplanetary medium \citep{soja2014probing,calves2014observations,kummamuru2023monitoring} and deep space navigation. A brief review of the latter VLBI applications in which spacecraft are observing targets is given in \cite{LIG+SSR-2023}. In most cases of VLBI observations of spacecraft within the Solar System, the distance to the target is smaller than the so-called Fraunhofer criterion \citep{Krish-Ram-2017}. In this case, it is appropriate to use the term near-field VLBI observation.

\begin{figure*}[h]
\centering
\includegraphics[width=1.0\textwidth]{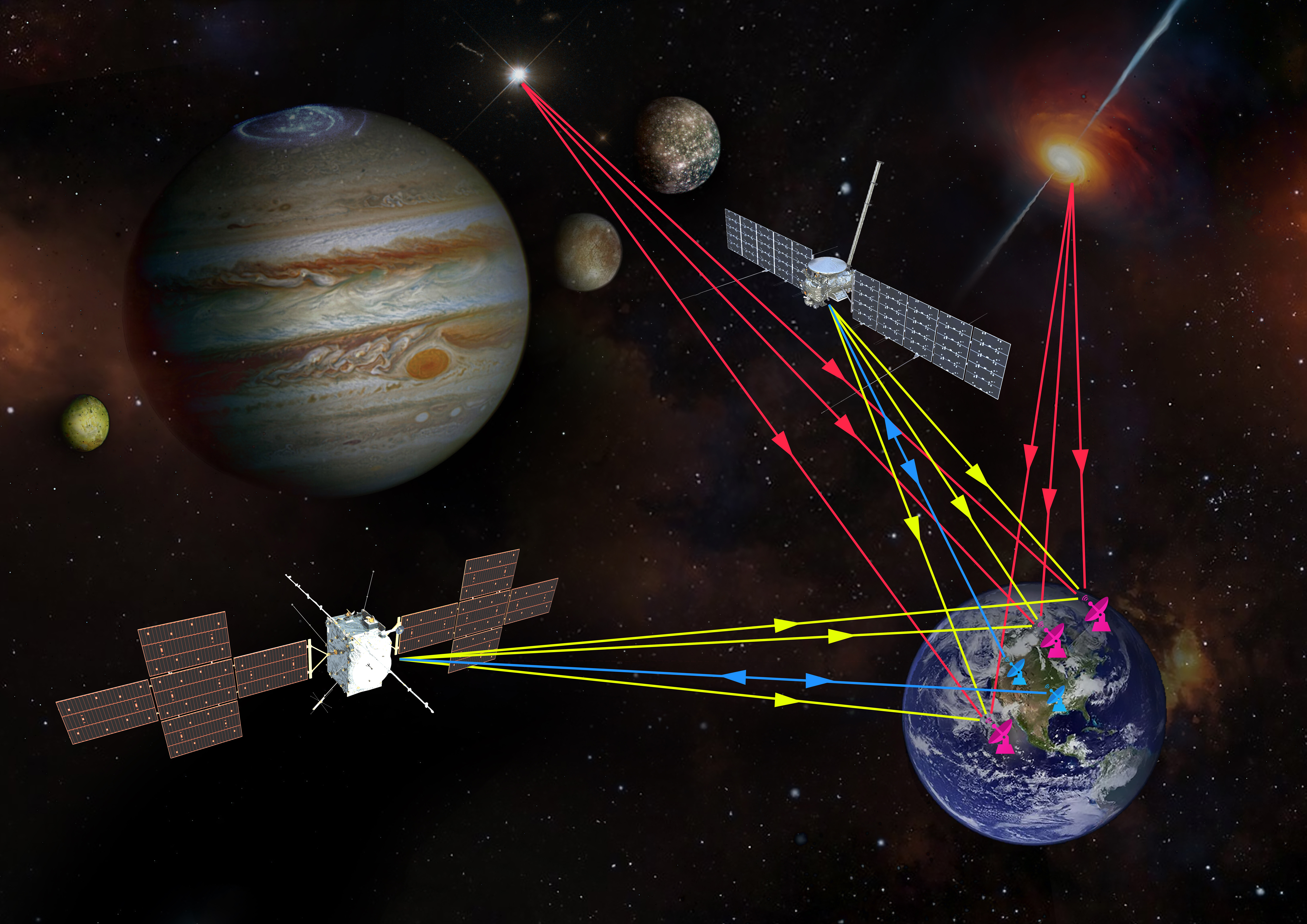}
\caption{A schematic configuration of PRIDE (not to scale) supporting two Jovian missions, JUICE (the lower left spacecraft) and Europa Clipper (the upper right spacecraft). The spacecraft are in two-way radio contact (blue arrows) with mission-specific Earth-based tracking stations (blue Earth-based antennas). Earth-based VLBI radio telescopes (three magenta Earth-based antennas) receive radio signals from the spacecraft (yellow arrows) and emission of distant natural celestial radio sources (red arrows). The configuration can operate at X- and/or K$_{a}$-bands, (8.4~GHz and/or 32~GHz, respectively).}
\label{fig:PRIDE_experiment}
\end{figure*}


The Planetary Radio Interferometry and Doppler Experiment \cite[PRIDE,][]{gurvits2013planetary,LIG+SSR-2023} utilizes data recording and processing technology developed for `traditional' astronomical VLBI but with the target source being a spacecraft (see  Fig.~\ref{fig:PRIDE_experiment}). The main measurables of PRIDE are the spacecraft's lateral celestial position and radial velocity. The former is provided by VLBI phase-referencing, the latter -- by estimating the Doppler shift of the spacecraft carrier tone. The PRIDE methodology is described in \cite{LIG+SSR-2023, duev2012spacecraft,bocanegra2018planetary}, and corresponding software implementation in \cite{molera2021high}. PRIDE has been selected by the European Space Agency (ESA) as one of the eleven science experiments of the Jupiter Icy Moons Explorer \cite[JUICE, ][and a special issue of Space Science Reviews 2023--2024, and references therein]{clavel2010cosmic, grasset2013jupiter, titov2015jupiter,fletcher2023ssr,Van_Hoolst+2024SSR}, the first ESA's Cosmic Vision 2015––2025 large-class flagship science mission. It was launched on 14 April 2023. 
JUICE will conduct 
investigations of Jupiter and its Galilean moons, particularly Ganymede, but also 
the other
moons Callisto and Europa. One of the scientific focuses of the mission is the potential habitability of these icy worlds. 

Jupiter in situ 
studies began in the 1970s with NASA's Pioneer program \citep{fimmel1980pioneer}, Voyager program \citep{heacock1980voyager, kohlhase1977voyager}, succeeded by the Galileo mission \citep{russell2012galileo, johnson1992space} in 1995--2003. NASA's Juno mission \citep{bolton2017topical} has been exploring the Jovian system since July 2016. Its goals are thorough studies on the planet's core and magnetic field, the composition of Jupiter's atmosphere, and the planet's auroras. NASA's Europa Clipper mission \citep{phillips2014europa,howell2020nasa} is expected to launch in 2024 and join JUICE for simultaneous in situ studies of the Jovian system from 2030. The Juno, JUICE and Europa Clipper missions are highly synergistic and will considerably deepen the understanding of Jupiter and its system.

The primary science objective of PRIDE for the JUICE mission is to support the improvement of the Jovian system ephemerides \citep{dirkx2017contribution,fayolle2022decoupled,fayolle2023}, which will enhance the science return for studies of the origin, evolution and potential habitability of the Galilean moons. The detailed descriptions of PRIDE--JUICE science objectives are provided in \citep{LIG+SSR-2023}.

\begin{table}[h]
\caption{The world VLBI arrays. The number of telescopes in the arrays and their X- and K$_{a}$-band capabilities are indicative and might vary with a tendency toward increase. Some telescopes might be members of more than one array. See \cite{LIG+SSR-2023} for more details.}\label{tab:vlbi-arrays}
\begin{tabular*}{\textwidth}{@{\extracolsep\fill}lcccccc}
\toprule%

VLBI Array & Number of 
telescopes & X-band & K$_{a}$-band \\
\midrule
European VLBI Network (EVN)  & 22 & 22 & 6 \\
U.S. Very Long Baseline Array (VLBA) & 10 & 10 & 10 \\
Japanese VLBI Network (JVN) & 9 & 9 & -- \\
Chinese VLBI Network (CVN) & $>$6 & 6 & 1 \\
Korean VLBI Network (KVN) & 3 & -- & 3 \\
East-Asia VLBI Network (EAVN) & 15 & 3 & -- \\
Australian Long Baseline Array (LBA) & 5 & 5 & 2 \\
University of Tasmania Array  & 5 & 5 & -- \\
\botrule
\end{tabular*}
\end{table}

\begin{figure*}[h]
    \centering
    \includegraphics[width=0.99\textwidth]{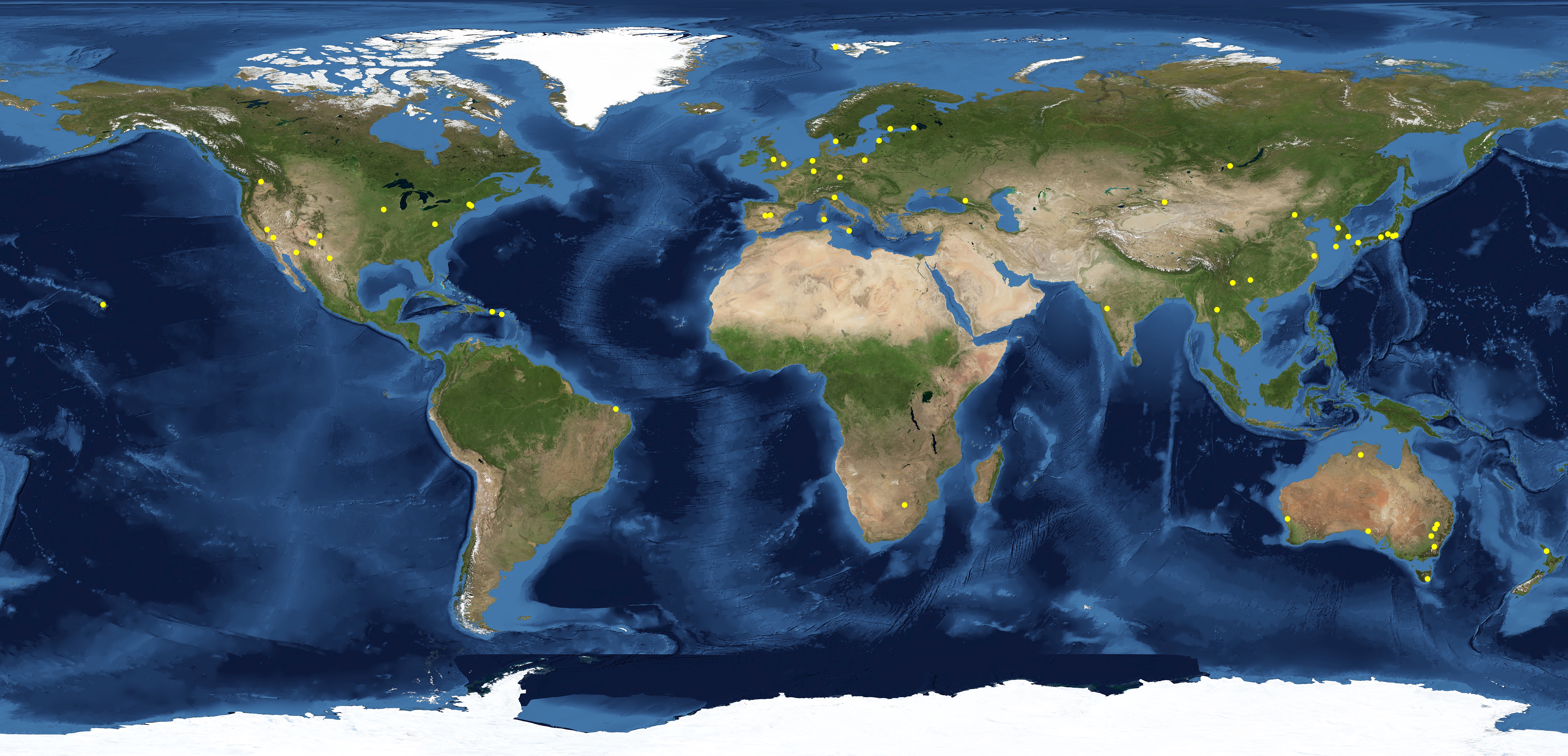}
    \caption{Global distribution of VLBI radio telescopes (shown by yellow dots). Most of the shown telescopes can operate at X-band (8.4~GHz), some at K$_{a}$-band (32~GHz). Several of the latter can do simultaneous X/K$_{a}$ observations. The map reflects the status as of March 2024. 
    During the operational lifetime of the JUICE mission, this map might change as more VLBI-compatible radio telescopes come online. Credit: the background map is a satellite image composite provided by Blue Marble Next Generation, courtesy of NASA Visible Earth (\url{https://visibleearth.nasa.gov}).
}
    \label{fig:VLBI-world}
\end{figure*}

In a typical PRIDE session, a network of Earth-based radio telescopes observes the JUICE spacecraft in phase-referencing VLBI mode. PRIDE does not have any dedicated onboard instrumentation and relies on the mission's radio communication systems. The main instrumental asset of PRIDE is the global network of VLBI-equipped radio telescopes, data transfer and data processing facilities. For a particular PRIDE observing run, a combination of Earth-based radio telescopes can be composed of a large set of facilities distributed over the globe, as shown in Fig.~\ref{fig:VLBI-world} and listed in Table~\ref{tab:vlbi-arrays} as of 
March 2024. In most cases, the VLBI telescopes are built and operated in the interests of galactic and extragalactic astronomy, astrometry and geodesy, and are unrelated directly to the JUICE or any other space science mission. PRIDE uses these resources in an ad hoc regime. The telescopes involved in PRIDE observations receive the radio signal transmitted by spacecraft. In PRIDE--JUICE, the nominal observing mode is at the X-band (8.4~GHz), but observations at the K$_{a}$-band (32~GHz) are possible too. 

The VLBI technique entered the domain of planetary science and exploration more than half a century ago -- see a review of planetary and space science missions supported by VLBI in \cite{LIG+SSR-2023}. In most cases, VLBI observations of planetary probes were conducted as ad hoc complements, not a part of nominal science suites of the mission (e.g., VLBI tracking of the ESA's Titan Huygens Probe \cite{pogrebenko2004vlbi,Lebreton+2005Nature}, and VLBA observations of the Cassini spacecraft \cite{Jones+2011AJ}). In some relatively rare cases, VLBI tracking was included as a nominal component of the mission's scientific operations. This was the case of the Venusian balloons of the VEGA mission and support to the Giotto Halley comet rendezvous \cite{VEGA-1986Venus, VEGA-RZS-1990, Pathfinder-1986}. The Chinese exploration programs of the Moon (Chang'e \cite{Changhe-R-2010S}) and Mars (Tianwen \cite{Tianwen-2022SSPMA}) include VLBI tracking as a nominal state vector determination and navigation technique. The Japanese Lunar Probe SELENE \cite{SELENE-VERA-2011} and Venusian solar sail IKAROS \cite{IKAROS-2011} also relied on VLBI support. However, in all these examples, VLBI tracking was exploited over a relatively short mission duration.

The VLBI component of the JUICE mission implemented as PRIDE differs from the above predecessors in the operational duration: 8.5 years of the cruise phase and at least 3 years of the Jovian tour. It is also the first occasion of the inclusion of PRIDE-VLBI as a nominal component in a mission to the outer Solar System. While the methodology of VLBI tracking of planetary missions has been developed and demonstrated by the PRIDE group over the past 20 years \citep[see][and references therein]{LIG+SSR-2023}, the higher load on VLBI operations as a nominal component of the JUICE mission required new approaches for planning and implementing a massive observing campaign using global VLBI networks. These ground-based VLBI operations should be properly integrated with the equally sophisticated but fundamentally different operational routines of the JUICE planetary mission. In this paper, we present this interface as well as our developments to prepare for the high operational load once in the Jovian system and give examples of consistent operational routines of the ground-based VLBI networks and the JUICE mission.

The paper is organized as follows: the 
configuration of PRIDE and its interaction with JUICE, as well as the science requirements and how it works in conjunction with other instruments, are covered in Section \ref{Imp}. In Section \ref{planning}, details on the experimental planning and scheduling methods are discussed. There, the various reasons why this will impact the data quality are described, and the various opportunities during the cruise phase where this impacts on the JUICE mission will be outlined. In Section \ref{software}, we present the specific tools that have been developed to automate planning and scheduling, show the application and results of these tools for the first PRIDE 
K$_{a}$-band observations of the BepiColombo spacecraft, and provide numerical simulations to quantify the potential impact that these tools could have on science return from PRIDE observations during the Venus flyby experiment. Finally, the concluding remarks and outlook are provided in Section \ref{summary}. 

\begin{figure}[h]
\centering
\includegraphics[scale=0.39]{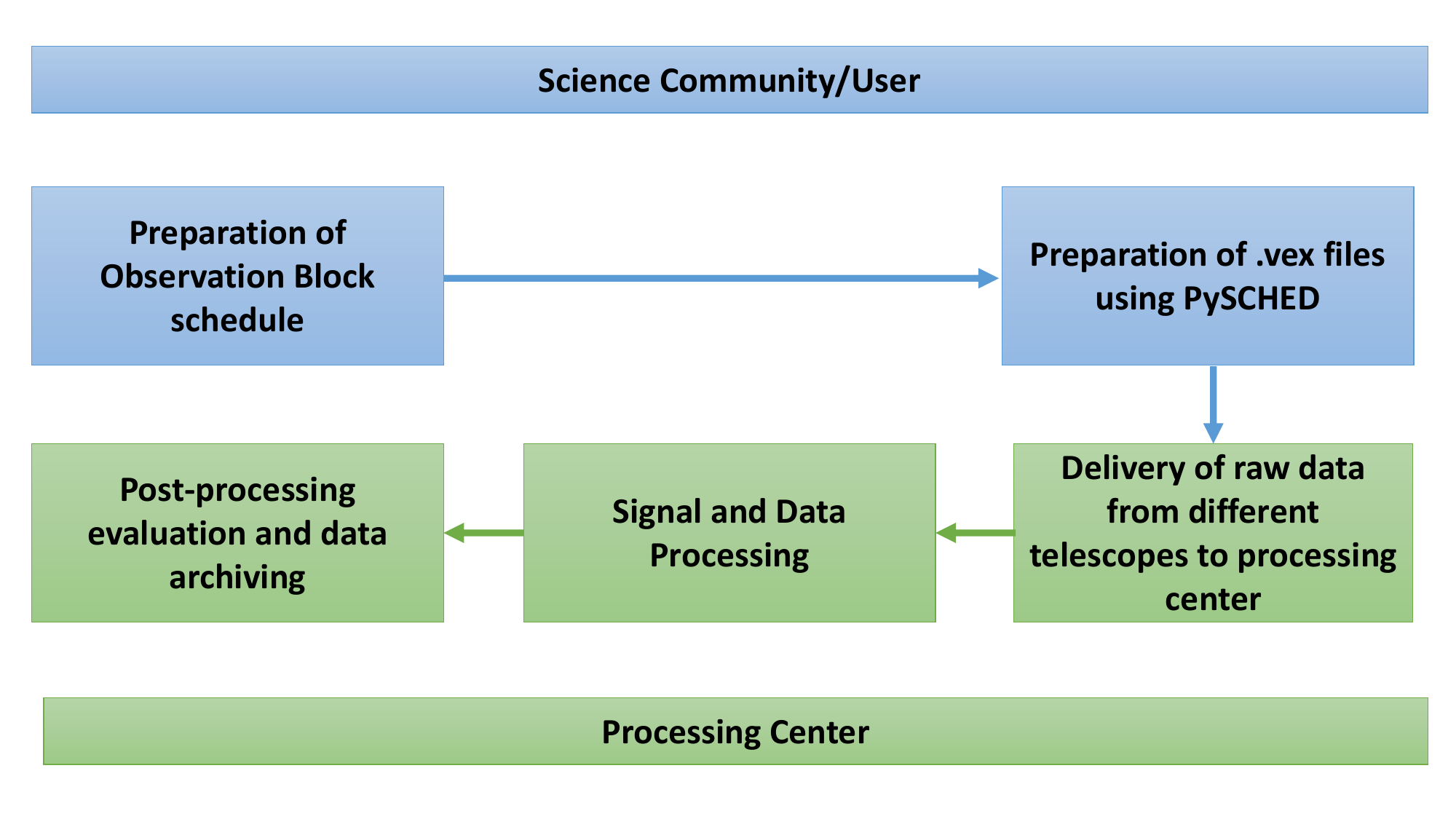}
\caption{PRIDE VLBI observation run activities and data processing. The main steps involve the scheduling of observation blocks, the preparation of .vex files, and the collection of calibration and baseband data for each telescope. The final visibility data released to the PI is bundled with calibration data. For near-field or far-field VLBI, signal processing can include delay models as well as optional processing like pulsar processing, space VLBI antenna support, multiple phase centers for wide-field VLBI, spectral windowing, or customized channelization. Data review and archiving to EVN or Planetary Data System (PDS) Standard is the final step.}
\label{fig:pride_activities}
\end{figure}

\section{PRIDE--JUICE basic configuration, requirements, and implementation}
\label{Imp}

On April 14 2023, JUICE was launched from the Guiana Space Centre in Kourou, French Guiana, using the Ariane 5 ECA (Evolved Cryogenic, model A) launch vehicle \citep{witasse2022juice}. The mission is comprised of three phases, namely, a cruise phase, a Jupiter tour beginning in July 2031 with the Jupiter Orbit Insertion (JOI), and a Ganymede phase beginning in December 2034 with the Ganymede Orbit Insertion (GOI), and ending nominally in September 2035. As per the most recent mission’s Consolidated Report on Mission Analysis (CReMA) 5.0b23 \citep{boutonnet2024juice}, JUICE will conduct numerous flyby studies of Ganymede, Callisto, and Europa before entering the polar orbit around Ganymede in 2034. The cruise phase, which will span for roughly 8 years prior to the JOI, will feature a flyby of the Moon and Earth in August 2024, a flyby of Venus in August 2025, and two Earth flybys in September 2026 and January 2029, respectively. As a scientific possibility, a flyby of one or more asteroids has been investigated \citep{witasse2021juice}.

We summarize the JUICE radio systems, the signal from which PRIDE uses to generate its observables, in Section \ref{radio}, followed by a brief overview of the PRIDE--JUICE science objectives and operations in Section \ref{scienceObjectives}. 

\subsection{JUICE radio system}
\label{radio}
The communications subsystem of the spacecraft is composed of a redundant set of Deep Space Transponders (DSTs) that use the X-band for the uplink and the X- and K$_{a}$-band for the downlink. The science payload also includes the KaT transponder, which is a part of the onboard radio science instrument 3GM \citep[Gravity and Geophysics of Jupiter and the Galilean Moons;][]{Iess-3GM-2013}. This transponder allows a K$_{a}$-band up- and downlink to be used for the generation of high-accuracy radio science Doppler and range data. The spacecraft antennas consist of a stationary high-gain antenna (HGA) (dual-band X and K$_{a}$), a steerable medium-gain antenna (MGA) (dual-band X and K$_{a}$), and two stationary X-band low-gain antennas (LGA). During the cruise, X-band uplink and downlink are the default settings.

The JUICE mission is operated by the ESA ground segment, consisting of the Mission Operations Centre (MOC) and the Science Operations Centre (SOC). MOC is in charge of ground segment development and spacecraft operations, while SOC is in charge of arranging, preparing, and sending the science operation requests to the MOC and coordinating the distribution of the data obtained from the MOC. The PRIDE ground segment can conduct observations whenever the spacecraft emits a radio signal for communication or radio science. 
Since the spacecraft already has onboard radio instrumentation required for PRIDE installed for other purposes (like navigation and communication), PRIDE does not affect the mission’s payload mass budget.

\subsection {PRIDE--JUICE Science Objectives and Operations }
\label{scienceObjectives}

Most of the science objectives of PRIDE are addressed by exploiting the natural synergy between PRIDE \citep{gurvits2013planetary} and 3GM \citep{Iess-3GM-2013}. 3GM will provide  highly precise range and
Doppler measurements, following procedures developed for the Mercury Orbiter Radio science Experiment (MORE) experiment on BepiColombo \cite[which is very similar to 3GM;][]{CappuccioEtAl2020b}. PRIDE will provide Doppler estimates at each receiving radio telescope in addition to the Doppler data obtained by 3GM at a dedicated tracking station. PRIDE can in principle operate with any radio signal coming from JUICE, but the typical operating mode for tracking will use the coherent downlink, from an uplink of an ESTRACK station. The noise parameters of PRIDE measurements are enhanced if the JUICE radio signal is not only transponded at X-band, but at Ka-band as well (when available, from a combined X- and Ka-band uplink), as this permits calibration of noise due to charged particles. This dual-frequency operation is facilitated by the 3GM Ka-band transponder (KaT). The operational advantage of PRIDE is in its ability to use multiple Earth-based receiving antennas thus having higher chance to pick up the best weather, ionosphere conditions as well as higher elevation of JUICE over local horizon (see an example of this advantage in VEX radio occultation experiments described in \cite{bocanegra2019venus}). For further discussion on the role of PRIDE in the overall JUICE radio science experiments see \cite{LIG+SSR-2023}. Furthermore and most importantly for the topics discussed here, PRIDE will provide VLBI data for the lateral spacecraft position in the International Celestial Reference Frame \citep[ICRF;][]{charlot2020third}. 



\begin{figure*}
    \centering
    \includegraphics[scale=0.48]
     {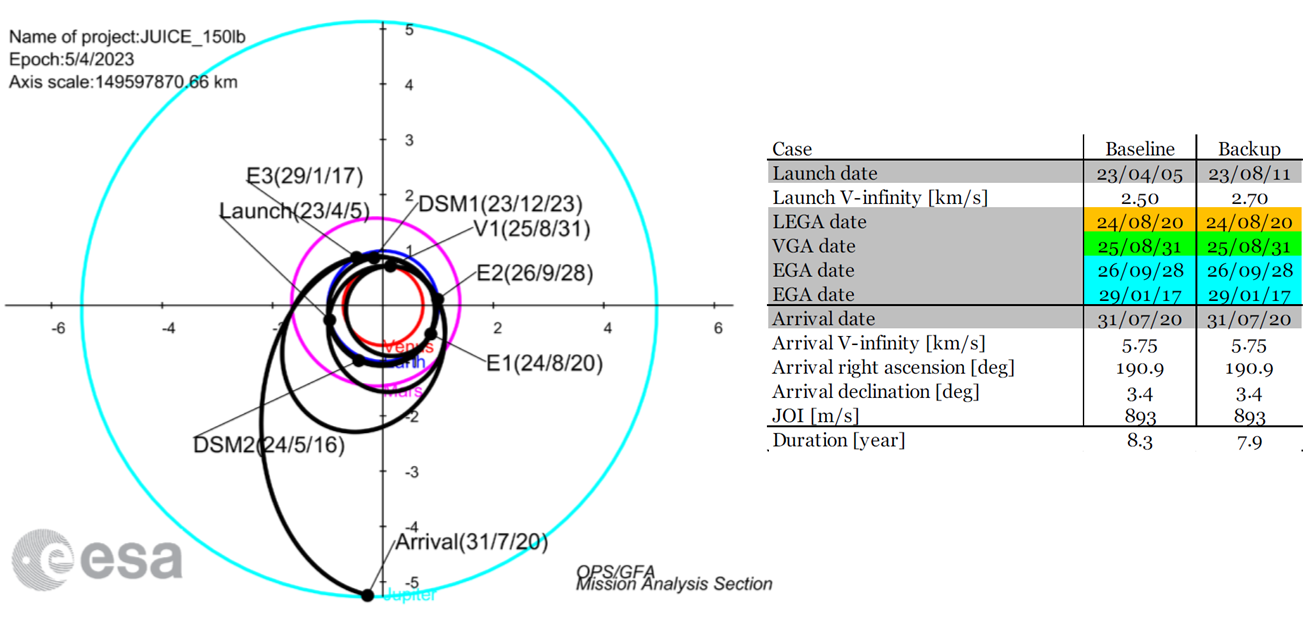}
    \caption{Interplanetary cruise phase and summary of transfer options for launch in 2023. Colour-code is used: green for Venus Gravity Assist (VGA), blue for the Earth Gravity Assist (EGA), orange for the Lunar-Earth Gravity Assist (LEGA) \citep{boutonnet2024juice}. Note that the JUICE launch occurred a little more than a week earlier than the baseline plan in this figure, which does not impact the general trajectory planning shown here.}
    \label{fig:cruise_phase}
\end{figure*}

PRIDE VLBI measurements are anticipated to improve the calculation of Jupiter's normal positions (out-of-plane direction) and of the Galilean satellites. particularly Callisto \citep{dirkx2017contribution,fayolle2024}.  In addition, the PRIDE VLBI data will be valuable as an independent validation of ephemeris solutions and, as such, will be of great value in developing a fully coupled dynamical solution of the JUICE spacecraft, the Galilean satellites, and Jupiter itself \citep{fayolle2022decoupled,fayolleEtAl2022}. A complete overview of the synergies with other instruments is discussed by \cite{LIG+SSR-2023}. In addition to science data products extracted from determining spacecraft dynamics, PRIDE will be used to perform radio occultation experiments on the Galilean satellites and Jupiter as probes for the atmospheres of these bodies, as done in the past for Venus using PRIDE \citep{bocanegra2019venus}. The very large collecting area of the radio telescopes involved in PRIDE, compared to regular single tracking station observations, as well as the more diverse observing geometry and local meteorology conditions facilitated by PRIDE's global nature, will allow facilitating a better quality occultation measurements. Using multiple stations for PRIDE signal detection could reduce propagation errors due to the Earth's atmosphere, improve spatial resolution and geometric accuracy, improve overall signal strength, and minimize noise. This increases the signal-to-noise ratio (SNR), which ensures higher-quality observations and is essential in detecting faint signals from spacecraft during occultations.
The study by \citep{bocanegra2019venus} demonstrated that occultations with Venus Express (VEX), utilizing PRIDE stations, achieve deeper penetration into the Venusian atmosphere compared to those performed with the dedicated Deep Space Network (DSN) and ESTRACK facilities.
PRIDE will also be useful in studying the interplanetary medium, as demonstrated in \cite{molera2017analysis}.

The operations of the ground segment of PRIDE solely depend on the ability and availability of ground-based radio telescopes to conduct PRIDE observations. The latter is defined by the telescope--spacecraft geometry configuration, the mission radio-link schedule, and the type of radio signal (carrier/modulated spectrum). In a nutshell, the PRIDE implementation solely uses existing technologies as a piggyback application during any interaction between the major Earth-based assets of PRIDE and the JUICE mission. However, compared to regular radio science for planetary spacecraft, planning PRIDE observations has the added complexity that the quality of all science products derived from the VLBI data will depend on the quality of the available phase calibrator(s) and the (time-variable) visible network of stations during the observation, which all have an influence on the quality of the angular position of JUICE as produced by PRIDE. These complexities are not present for `regular' radio science observations, and their impact is therefore not studied in much detail. The desire to automate and (where possible) optimize for these added complications in large part motivates the need for the work presented here.

\begin{table*}[ht]
\centering
\caption{Cruise phase flyby summary according to \citep{boutonnet2024juice}.}
\begin{tabular}{c|ccccccc}
\hline
\textbf{Flybys} & \textbf{Time UTC} & \textbf{Altitude} 
\\ 
 & & [km] 
 \\ \hline

Moon\textunderscore FB & 2024-08-19T19:34:20 & 750.11 \\
Earth\textunderscore FB & 2024-08-20T20:15:27 & 6402.95 \\
Venus\textunderscore FB & 2025-08-31T06:18:12 & 5081.73 \\
Earth\textunderscore FB & 2026-09-28T11:53:56 & 8641.88\\
Earth\textunderscore FB & 2029-01-17T18:27:20 & 4643.97 \\
\end{tabular}
\label{tab:cruisephase}
\end{table*}

\begin{sidewaystable}
\centering
\caption{The types of PRIDE observations during the cruise phase.Each experiment's duration is $\ge 2$~h}.
\begin{tabular}{c|p{8cm}p{3cm}c}
\hline
\textbf{Experiment Code} & \textbf{Description} & \textbf{Experiment Type}  \\ \hline
PRIDE-EXP1	& Observations during communication/downlink sessions and/or Delta DOR (Differential One-way Range) sessions outside of flybys	& Nominal	  \\
PRIDE-EXP2	& Observations close to the Venus flyby. Acquisition of VLBI data for improvement of solar system ephemerides & Nominal  \\
PRIDE-EXP3	& Acquisition of occultation profiles at Venus and/or Moon (during Lunar flyby) & Occultations \newline (for science \newline and/or tests) \\

PRIDE-EXP4	& Acquisition of training data sets for operations in the Jovian system (`regular' mode, `in-beam with calibrator' mode, `in-beam with second spacecraft' mode) & Test and training  \\ 
\hline

\end{tabular}
\label{tab:experiments}
\end{sidewaystable}

\section{PRIDE--JUICE planning and scheduling}
\label{planning}

An overview of PRIDE experiment planning and scheduling methods for the various mission phases is presented in this section, as well as an overview of their application during the cruise phase. Note that, in the present context, the term `planning' is used for the general overall procedure of deciding when to do the observations, while `scheduling' is used for the more detailed procedure of deciding which ground stations, frequencies, phase calibrators, etc. to use. Many criteria must be considered before a PRIDE observation can be conducted, including the spacecraft transmission window, (in-beam) phase-referencing opportunities with nearby calibrators, and ground station availability. Planning an observation is significantly impacted by these considerations. This section outlines the steps for scheduling, planning, and preparing for a PRIDE observation.

First, the impact of planning and scheduling on PRIDE VLBI data quality is discussed in Section~\ref{develop}. A basic explanation of operational procedures in planning and scheduling is provided in Section~\ref{operationalProcedures}. 
The PRIDE operations for the cruise phase are covered in Section~\ref{cruise}, highlighting opportunities where our experiments for methodological validation will also be applied for scientific purposes. In particular, it is discussed how our planning and scheduling tools could be used to improve the science return from these experiments, in preparation for maximizing the contribution to the mission of extended PRIDE operations in the Jovian system.  

\subsection{Impact on data quality}
\label{develop}

The quality of the PRIDE science products derived from VLBI data relies on the observations of a nearby phase calibrator source \citep{rioja2020precise}, which is often a quasar with a well-defined position in the ICRF. Both spacecraft and calibrators must be viewed through the same patch of the sky to remove atmospheric interference due to the ionosphere and the troposphere, and instrumental phase errors. If a calibrator is too far from the target, variations in the atmosphere make it impossible to correct these effects. Usually, if the calibrator is within $\sim 2^{\circ}$ \citep{bocanegra2019venus} of the target, calibration can be performed. However, the telescopes must be steered back and forth between the target and the calibrator on a frequent basis, a procedure known as nodding. In cases where suitable reference sources are available and with proper planning and scheduling, calibration can be improved using in-beam phase referencing. This technique involves observing the reference source within the same telescope's primary beam as the target source, removing the need for nodding and improving data quality.


The accuracy and precision of the resulting VLBI data point of JUICE will depend on the quality and closeness of the phase reference source. In case suitable phase-reference sources close to JUICE during a period of specific scientific interest are not available, a dedicated observation campaign can be conducted specifically to densify the catalogue of suitable reference sources in the patch of the sky of interest. However, such a campaign must be executed separately from and well in advance of 
the spacecraft VLBI observations and should only be done if the ultimate science case for doing so is sound. An example of such a preparatory VLBI experiment is described in Section~\ref{venus-flyby} for the case of the JUICE Venus flyby.

The PRIDE technique can also be applied to simultaneously receive the signals from two or more transmitting spacecraft, which can be used to create multi-spacecraft in-beam VLBI observations, providing the relative angular position of the two spacecraft in ICRF to be determined with an accuracy similar to regular VLBI observables. In the context of JUICE, this will provide unique opportunities for conducting concurrent in-beam PRIDE observations of JUICE and NASA's Europa Clipper, with potentially significant advantages in creating a consistent and coupled dynamical solution for the Galilean satellites from the data sets of the two missions \citep{fayolleEtAl2022}. { Using the ExoMars Trace Gas Orbiter \citep{gibney2016mars} and Mars Express \citep{schmidt2003mars}, a preliminary study of such a dual-spacecraft PRIDE experiment was carried out by \citep{klindvzic2018planetary} in Mars orbit.} A comparison of the propagation effects on simultaneous tracking of Mars Express and Tianwen-1 was reported by~\citep{masdiana}. Provided that the opportunity arises, a future PRIDE observation campaign during the cruise phase will be planned where a second spacecraft is visible within $\sim 2^{\circ}$ of JUICE in preparation for the operations in the Jovian system.


\subsection{Operational procedures}
\label{operationalProcedures}
The overall and nominal activities of the PRIDE observation run are shown in Fig.~\ref{fig:pride_activities}. It depicts how PRIDE observations are planned, including which radio telescopes will observe which spacecraft at what time. The observation is planned using custom-built planning software. The scheduling is accomplished by 
VLBI specialized software called 
SCHED or, more recently, pySCHED\footnote{https://github.com/jive-vlbi/sched}. The input for the scheduling software is created from the results of the developed planning tool and ad hoc scripts\footnote{https://gitlab.com/gofrito/makekey} for calculating spacecraft coordinates and adding a variety of desired scheduling parameters. The accurate computation of source visibilities, taking into account each antenna's sensitivity, antenna mounting, slewing rates, and source strength are essential components of a scheduling tool. The outputs are standard-format schedules that each participating radio telescope uses to acquire its data. The antenna steering, signal chains, and sampler (digitizer) configuration, as well as the recording mode, are all contained in the station-specific schedule file, which is also defined in the process of scheduling. 

A proper time schedule, including uplink and downlink sessions, UT range, and the predicted position of a spacecraft at different epochs, is required to observe spacecraft signals with VLBI radio telescopes. These inputs are anticipated from the SOC and/or MOC well in advance and should be prepared before an observation, according to the tracking session information and CReMA documentation \citep{boutonnet2024juice}. After the observation planning and scheduling, each of the participating ground stations sends its raw data to JIVE in the prescribed format. The open-loop VLBI/Doppler signal processing methods described in \citep{duev2012spacecraft, molera2021high} are used to analyze this time domain received data. Finally, the mission's post-processing and archiving will be carried out in accordance with the Planetary Data System (PDS) archiving guidelines.

\subsection{Mission scenario and PRIDE operations during cruise phase}
\label{cruise}

\begin{figure}
    \centering
    \includegraphics[scale=0.39]
    {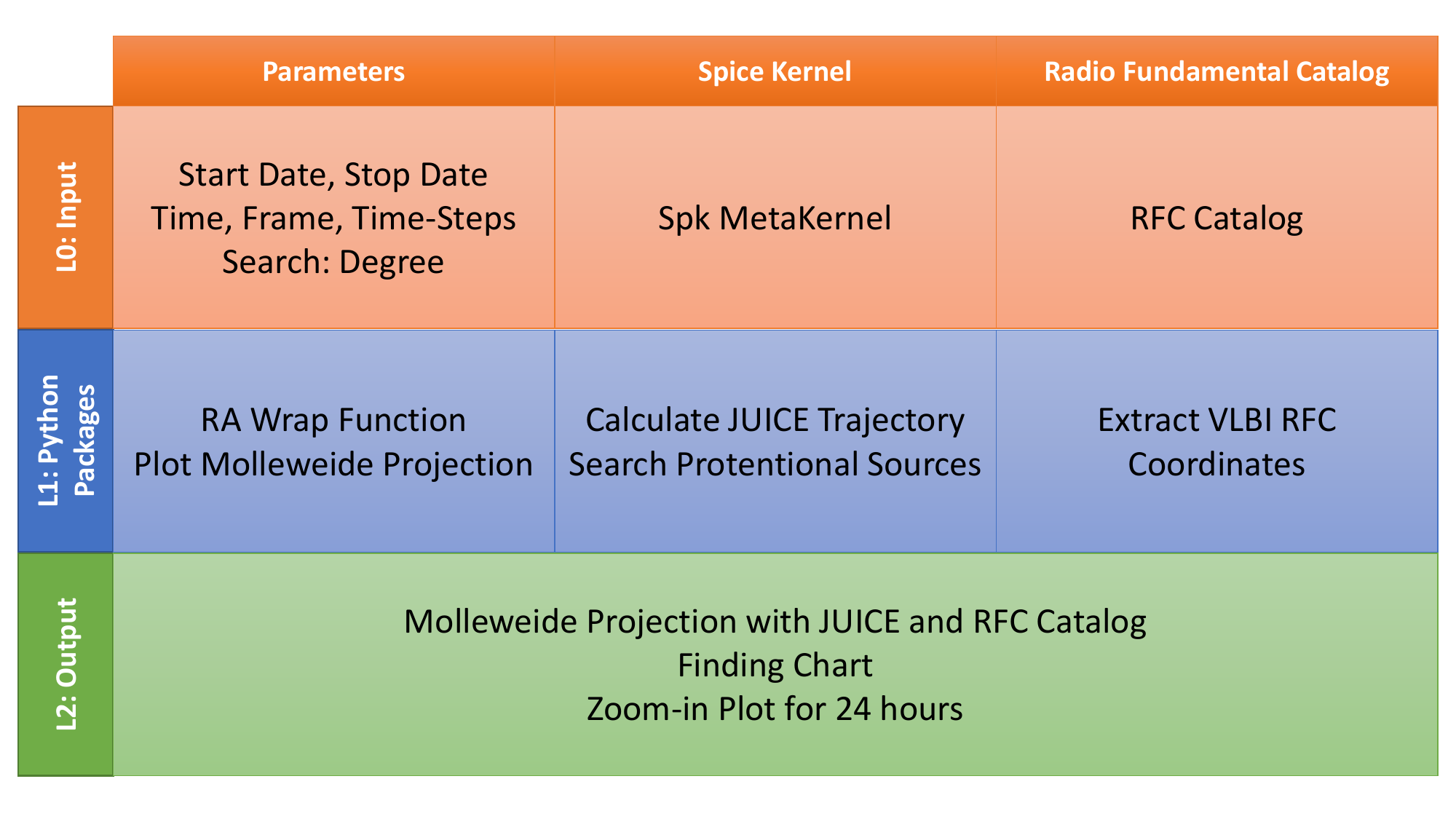}
    \caption{Software architecture for the PRIDE planning and scheduling tool. The first section of the software, shown in orange blocks, is employed as the input section and consists of user input, the Spice kernel, and the Radio Fundamental Catalogue (RFC). The second section of the program includes a specially created Python module, a Spicepy wrapper, and Astropy, which calculates the trajectory of the spacecraft (in light blue). The finding chart using Mollweide projection and zoomed-in plots (e.g., as shown in Fig.~\ref{fig:Planning_chart}) are produced using this software's third and output sections (in green).}
\label{fig:sw_arch}
\end{figure}

The experiment configuration described here is based on the JUICE Interplanetary Transfer Phase (ITP)/cruise phase, which runs from the end of the Near-Earth Commissioning Phase (NECP) up to six months prior to the execution of the Jupiter Orbit Insertion (JOI). The interplanetary transfers described in \citep{witasse2021juice,boutonnet2024juice} are the basis for this case study and are shown in Fig.~\ref{fig:cruise_phase}. The comprehensive list of planetary flyby events during the ITP period is given in Table~\ref{tab:cruisephase}. The transfer involves flybys of Venus, Earth, and the Moon in various geometric configurations, according to the CReMA  \citep{boutonnet2024juice} version 5.0b23. 

The PRIDE data processing pipelines and methodologies outlined in previous sections will be carried out in `live' tests during the mission's cruise phase in preparation for scientific operations in the Jovian system, in particular, to develop and test improved automation and data analysis procedures. Although these experiments are primarily carried out for testing and calibration purposes, the PRIDE data acquired during these sessions can also be used for a number of scientific applications, which are listed in Table~\ref{tab:experiments}.

ESA's deep space stations ESTRACK will be used to conduct spacecraft tracking throughout the interplanetary phase, primarily for navigation purposes. PRIDE will run its observations (PRIDE-EXP1, Table~\ref{tab:experiments}) in parallel with part of MOC's standard weekly orbit determination sessions during the ITP and nominal phases. PRIDE will undertake four to six spacecraft observing sessions each year for testing purposes. 

PRIDE is planning to do observations close to the Venus flyby (PRIDE-EXP2, Table~\ref{tab:experiments}). Of the entire cruise phase, the Venus flyby most closely mimics the situation of prime scientific interest for PRIDE: the flybys at the Galilean moons. During the Venus flyby, the spacecraft constraints are high due to thermal conditions (no nadir pointing in particular) and there will be severe limitations on payload operations \citep{boutonnet2024juice}. Provided that the MGA is on during the flyby (or very near it), the training experiment during the flyby will also provide VLBI data to improve the Venus ephemeris. Consequently, the data will also allow us to test our data analysis chain up to and including spacecraft orbit determination and flyby target normal point estimation \citep{fayolle2022decoupled}. As a showcase, Section \ref{tudat} summarizes the results of numerical simulations of PRIDE observations during the Venus flyby to quantify and highlight the potential impact that VLBI data quality, and therefore proper scheduling and preparation of the experiment, can have on the science return. 

In addition to the VLBI data acquired during the Venus flyby, the Doppler data acquired during ingress and egress of the planet will be useful for occultation studies of Venus' atmosphere (PRIDE-EXP3, Table~\ref{tab:experiments}).  An analysis of such data sets using Venus Express was presented by
\citep{bocanegra2019venus}, who provided a detailed model for the use of PRIDE in radio occultation measurements and its possibilities for Venus in particular. The Venusian atmosphere makes occultation measurements at Venus fundamentally different than those that could be acquired for the Galilean satellites, although the Venus occultation data could provide useful training data to prepare for PRIDE occultations of the Jovian atmosphere. Possible acquisition of PRIDE Doppler data during the lunar flyby could be used as a more representative test case for radio occultation and/or bistatic radar studies for later observations at the Galilean moons.

\begin{figure}
\centering
\includegraphics[scale=0.41]{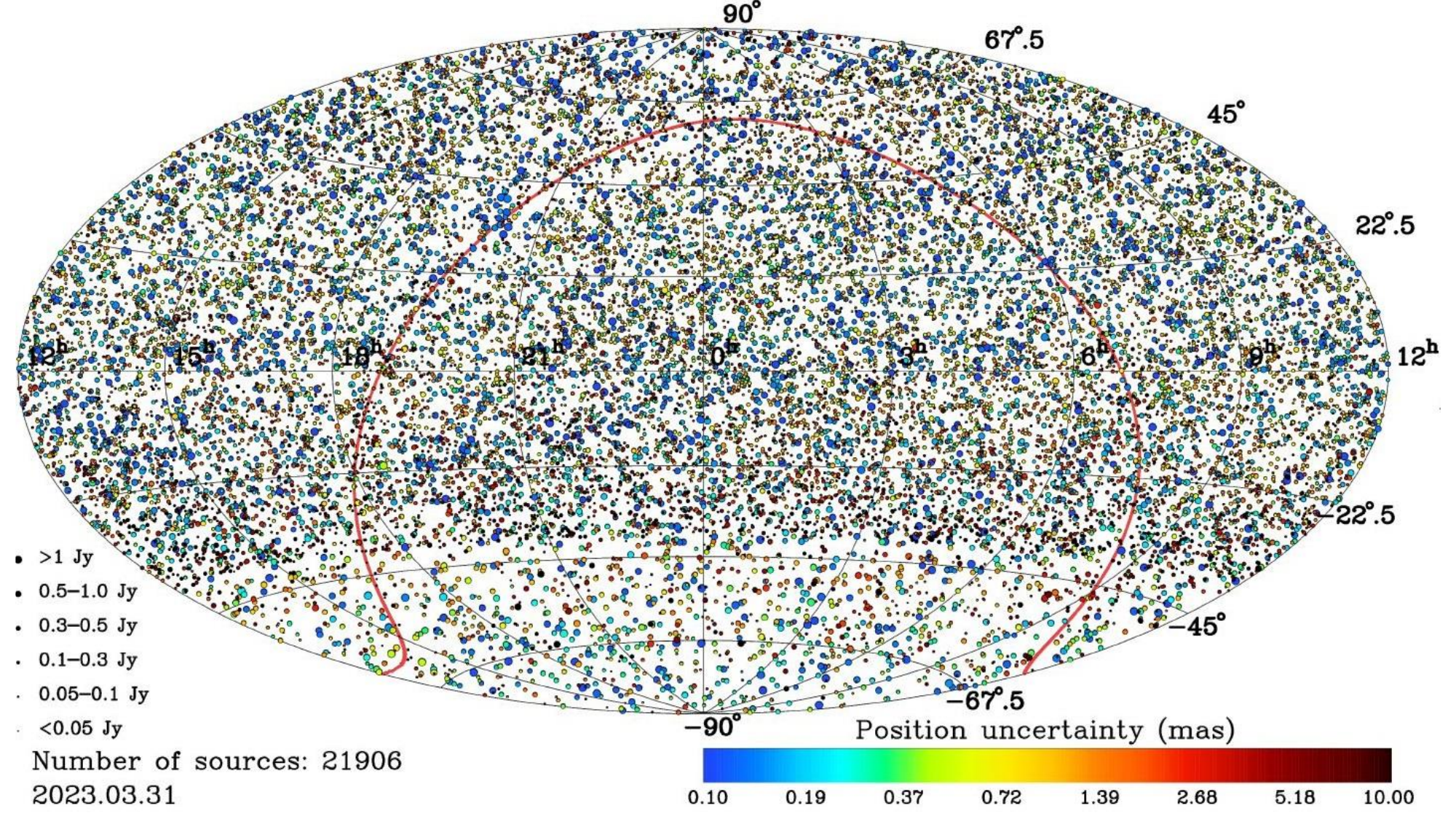}
\caption{Sky distribution of the total of 21,906 compact radio sources from the RFC catalogue}  \citep{PetrovL2022}. 
\label{fig:petrov_chart}
\end{figure}

Training data sets (PRIDE-EXP4, Table~\ref{tab:experiments}) obtained with an in-beam mode will be acquired if feasible. Here, we distinguish two types of observations: one where the signal from JUICE and a second spacecraft are tracked simultaneously, and one where the signal from JUICE and the phase calibrator are observed simultaneously (e.g., without nodding). The `in-beam with second spacecraft' mode will be most useful for testing and optimizing PRIDE operational and data analysis systems in advance of simultaneous observations of JUICE and Europa Clipper \citep{fayolleEtAl2022}. Whether the opportunity for such observations will arise is not yet clear (e.g. whether a second spacecraft will be present at a suitable geometry), and will be the topic of a dedicated analysis by the PRIDE team. In addition, opportunities during the `regular' training sessions (EXP1) where a suitable in-beam phase calibrator is available will be used for scheduling and data analysis procedures for in-beam observations.

\begin{figure}
\centering
\includegraphics[scale=0.26]{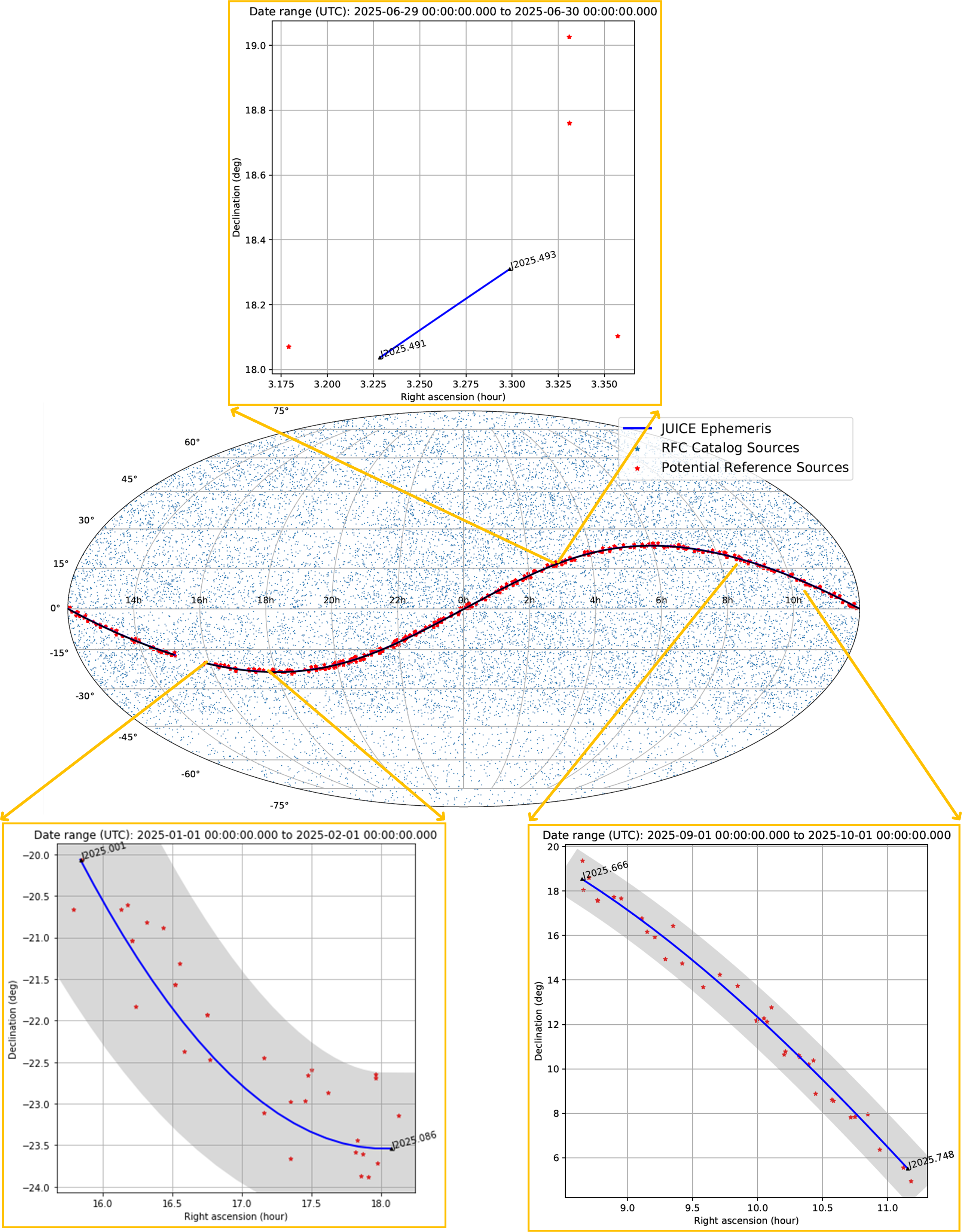}
\caption{PRIDE observational planning chart for the cruise phase of the JUICE mission (CReMA 5.0) for the time period of 2025.01.01 to 2025.12.31. The middle panel shows the whole sky in the Mollweide projection populated with the RFC celestial sources (blue dots). The upper and two lower panels (framed in orange color) show zoomed-in finding charts for periods of one day and one month, respectively. Red dots indicate potential reference sources from the RFC \citep{PetrovL2022} that are within $1^{\circ}$ of the JUICE spacecraft's ephemeris. Blue lines indicate the JUICE celestial track as per CReMA 5.0. The shadowed area indicates the celestial positions within $1^{\circ}$ from the JUICE celestial track. In general, the calibrators are distributed uniformly along the JUICE spacecraft sky track.}
\label{fig:Planning_chart}
\end{figure}

Finally, the interplanetary plasma scintillation (IPS) \citep{molera2014observations, molera2017analysis} of the spacecraft radio signal on the solar wind plasma will be accurately evaluated using all cruise phase test data sets. Such observations were first made in 2010 with ESA Venus Express \citep{molera2017analysis}, continued with Rosetta, and continue up until the present with numerous ESA/ NASA missions \citep{molera2021high}. In the last ten years, thousands of sessions have been undertaken using numerous VLBI radio telescopes. Additionally, an observational campaign has been launched to monitor solar activity, including coronal mass ejections (CME), by initiating radio observations when these events pass between the spacecraft and ground antennas and the JUICE communications system is active. It is anticipated that the signal JUICE transmits during its cruise phase will supplement the most recent observations. 

In addition to the objectives listed above, PRIDE can also perform short-notice observations under a special protocol established jointly by the PRIDE team and SOC as an exception or contingency. {When there is an urgent need for data collection and the typical observation planning approach is insufficient, these observational modes are activated.} This contingency operational mode is predicated on the fact that some of PRIDE's most important operational assets – Earth-based radio telescopes— are extremely sensitive and geographically scattered over the whole range of longitudes. PRIDE can deliver single-dish (signal detection, including Doppler measurements) or VLBI observations in this operational mode. In exceptional circumstances, the former can be carried out with only around 24 hours' notice.

\begin{figure*}
    \centering
    \includegraphics[scale=0.34]
    {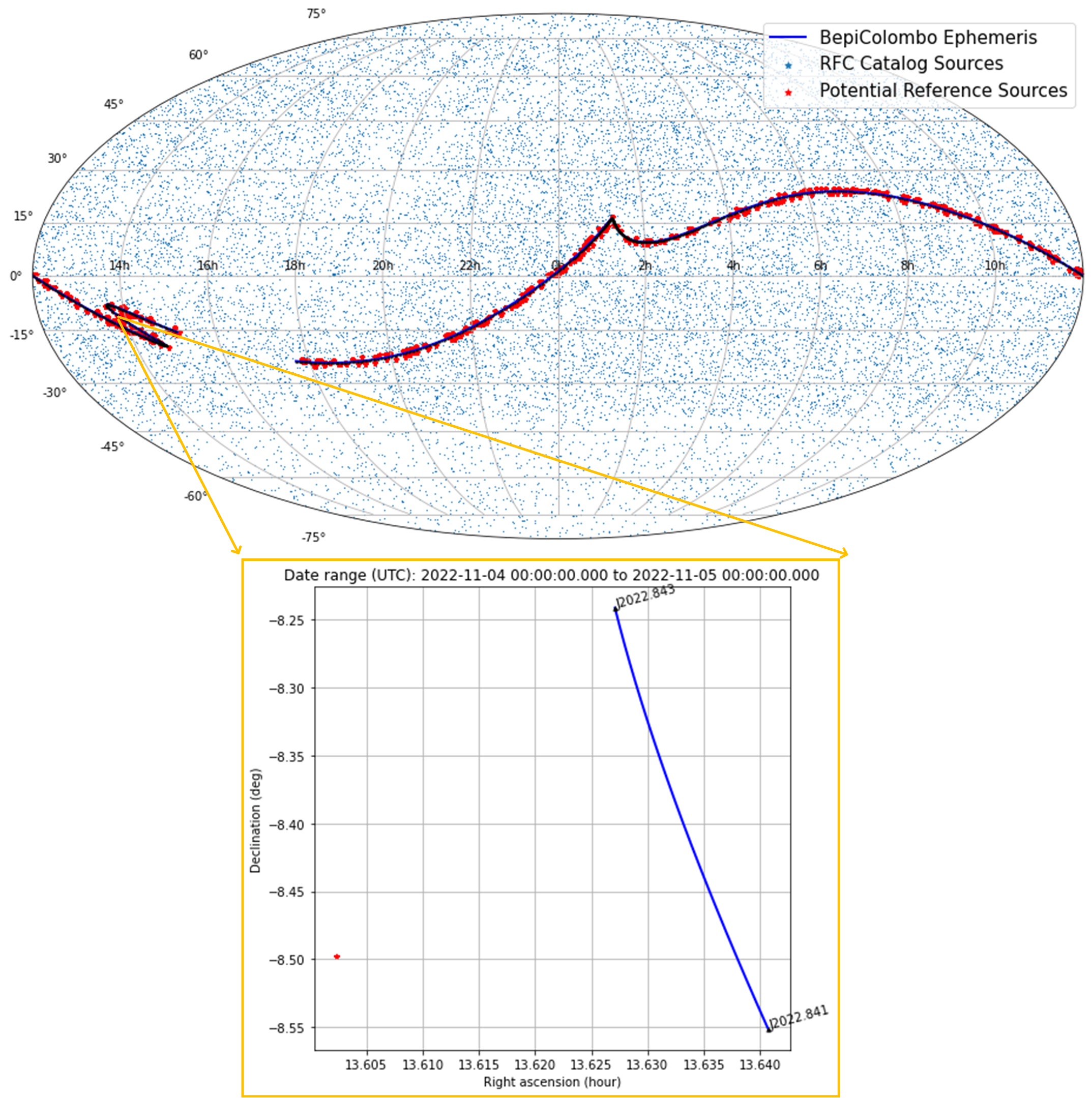}
    \caption{BepiColombo-PRIDE planning chart for January 1, 2022, to December 31, 2022. The upper panel shows the whole sky in the Mollweide projection populated with the RFC celestial sources (blue dots). The lower panel (framed in orange color) shows a zoomed-in finding chart for a period of one day. Red dots indicate potential reference sources from the RFC that are within $1^{\circ}$ of the BepiColombo spacecraft's ephemeris. Blue lines indicate the BepiColombo celestial track, along with any probable reference sources.}
    \
    \label{fig:Bepicolombo_PRIDE}
\end{figure*}

\section{Application to current and future observations}
\label{software}

 As described in Section~\ref{develop}, the initial step of the PRIDE implementation process involves identifying science opportunities at distinct mission phases in accordance with the mission's overall science operations plan. In this section, we present the tools we have developed to support the planning and scheduling and provide illustrative results for past and upcoming experiments. The specifics of the custom-developed observational planning tool, primarily the identification of potential phase referencing calibrators near the celestial position of the spacecraft at a given epoch, is discussed in Section~\ref{result1}. 
Section~\ref{bepi} describes the application of the PRIDE observation planning and scheduling tool to the BepiColombo mission, which represented the PRIDE team's initial acquisition of a spacecraft signal at K$_{a}$-band and demonstrated PRIDE's conceptual feasibility in K$_{a}$-band.
The planning of a preparatory VLBI experiment to characterise calibrator sources and to look for potentially suitable in-beam phase-reference calibrators prior to the JUICE Venus flyby in 2025 is presented in Section~\ref{venus-flyby}. Finally, Section~\ref{tudat} presents the results of numerical simulations to highlight the importance of effective planning and scheduling and proper selection of different phase calibrators for PRIDE--JUICE during its first end-to-end test for the acquisition of VLBI data and improving ephemerides: the Venus flyby (see Sections~\ref{cruise} \vp{}{and \ref{venus-flyby}}). 

\subsection{Planning and scheduling tools}
\label{result1}

For PRIDE experiments with {past planetary missions} such as Mars Express and Venus Express, numerous manual actions were required to meticulously prepare the necessary observation plans, scheduling, and related interchange files. The issue in the case of the VLBI spacecraft observations was that the widely used geodetic scheduling systems did not normally support spacecraft as radio sources.{ 
In addition, because spacecraft missions are dynamic and require scheduling and coordination not only between the receiving telescopes but also between the transmitting station and spacecraft, the procedure of scheduling observations presents a number of additional difficulties. Spacecraft orbit and visibility, spacecraft operational constraints, flexibility for real-time adjustments, coordination of multiple observing modes, distribution of baseline lengths, and data prioritization are the main challenges when scheduling VLBI observations of spacecraft.} 

To address these challenges and streamline the process as much as possible, we have created a specialized spacecraft observation planning and scheduling module. It makes it easy and practical to create realistic satellite observation plans for upcoming experiments {with only the bare minimum of required input from users}. Figure~\ref{fig:sw_arch} depicts the PRIDE planning and scheduling program's software stack. 

\begin{itemize}
  \item L0 (Input): This module contains all the experiment's input, such as the start and stop dates, reference frame, reference catalogue, SPICE metakernel, the time step, \textit{etc.}
  \item L1: with this module, potential reference sources are identified and classified based on the spacecraft trajectory. This module calculates the JUICE trajectory using Spicepy \citep{costa2018spice} (a Python wrapper for the SPICE toolkit \citep{acton1996ancillary}) and extracts the VLBI RFC coordinates using Astropy \citep{robitaille2013streicher}.
  \item L2 (Output): visualization in the form of a finding chart in Mollweide projection and zoom-in charts for a specific event are produced for the specific experiment of interest. 
\end{itemize}

The Radio Fundamental Catalogue (RFC) by \cite{PetrovL2022} is the most complete catalogue of compact radio sources currently available for {potential} calibrators' searches. The RFC was created by analysis of all available VLBI data produced under absolute astrometry and geodesy programs as well as other VLBI studies to offer milliarcsecond-accurate locations, maps, and estimates of correlated flux densities for thousands of compact radio sources. It includes a total of 21,906 sources detected (Fig.~\ref{fig:petrov_chart}) with VLBI 
between 1980 and 2023.

Making a so-called finding chart (as in Fig.~\ref{fig:Planning_chart}), which displays the target spacecraft's (JUICE) coordinates as seen from Earth over a period of time, is the first step of the tool. For the purpose of observing weak objects or conducting astrometry studies, phase-referencing calibrator sources are required. These sources must be compact, with a strong signal on all baselines (above 0.3 Jy), ensuring accurate phase calibration. The availability of appropriate stations on the ground and knowledge of stable radio sources distributed over the sky is critical for this. The second step involves creating a defined search strip of the sky centred on the spacecraft's ephemeris, and searching for radio sources inside this search strip. All these sources could be used as prospective PRIDE experiment calibrators and the proper selection of calibrators has an impact on the VLBI data quality. Figure~\ref{fig:Planning_chart} depicts the PRIDE observation planning chart for CReMA 5.0 during the cruise phase of the JUICE mission. These results are utilized to detect instances of in-beam phase referencing observation \citep{bocanegra2019venus} possibilities or, in general, PRIDE observing opportunities.

\begin{figure*}
    \centering
    \includegraphics[scale=0.46]
    {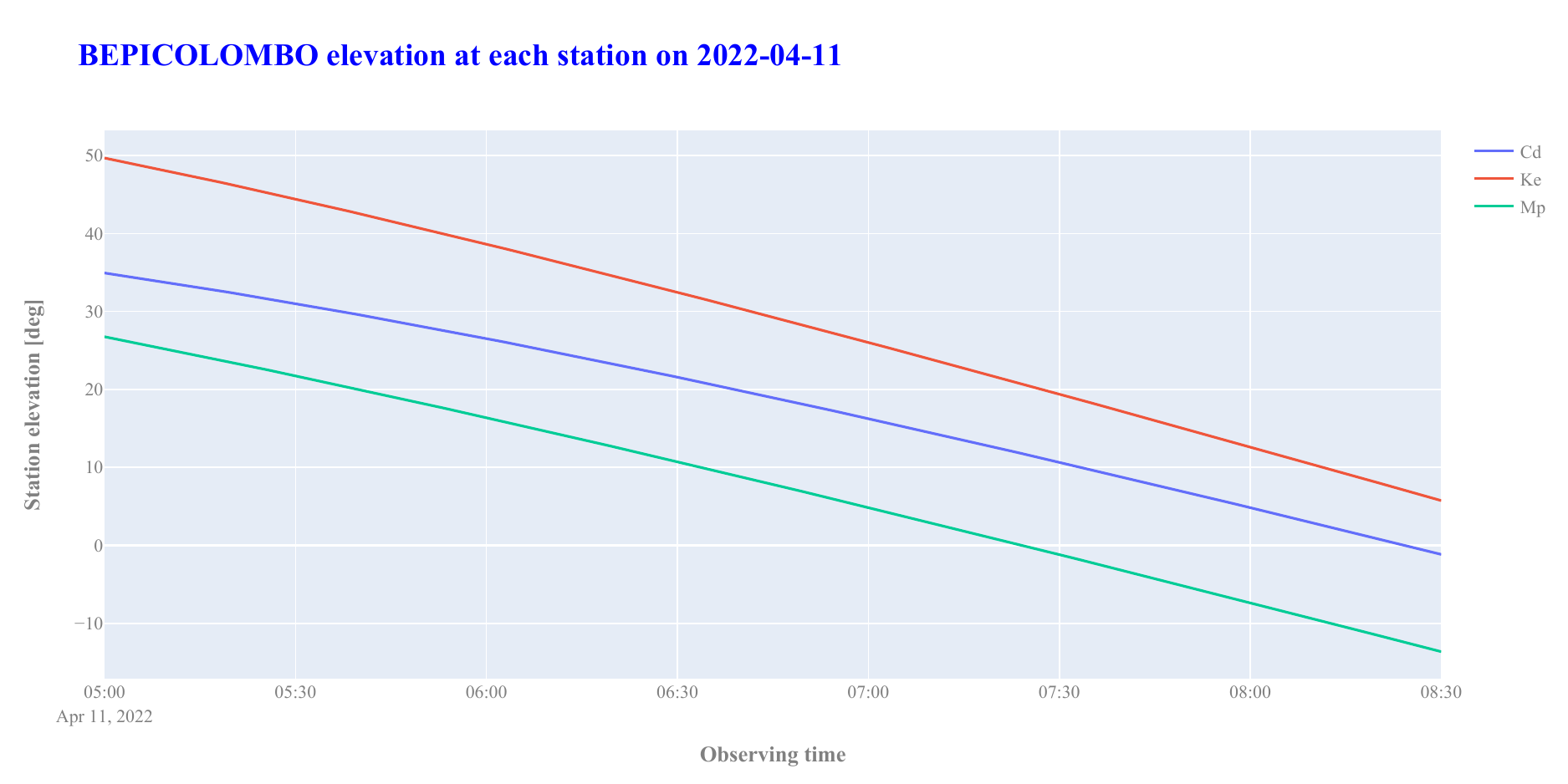}
    \caption{{Elevation visibility of the} three telescopes, showing elevation against UTC (uptime per telescope). Periods of observation (UTC) for the telescopes that took part in the phase-referencing VLBI section. The colors green, blue, and red represent the Australian stations Mopra (Mp), Ceduna (Cd), and Katherine (Ke), respectively.}
    \label{fig:ScVisibility_20220411}
\end{figure*}

 The PRIDE JUICE project relies heavily on the ecliptic plane densification of these calibrator sources, especially in  K$_{a}$-band (32 GHz). At present, the density of these sources along the ecliptic plane is insufficient for high-accuracy VLBI spacecraft tracking, according to the current list of phase calibrators as shown in Fig.~\ref{fig:petrov_chart} \citep{PetrovL2022}. Before arrival in the Jovian system, we will perform dedicated observational campaigns to create a dense grid of calibrator sources around Jupiter's orbit in order to achieve sub-nano radians accuracy in the spacecraft's lateral positioning. 

 \begin{figure*}
    \centering
    \includegraphics[scale=0.45]
    {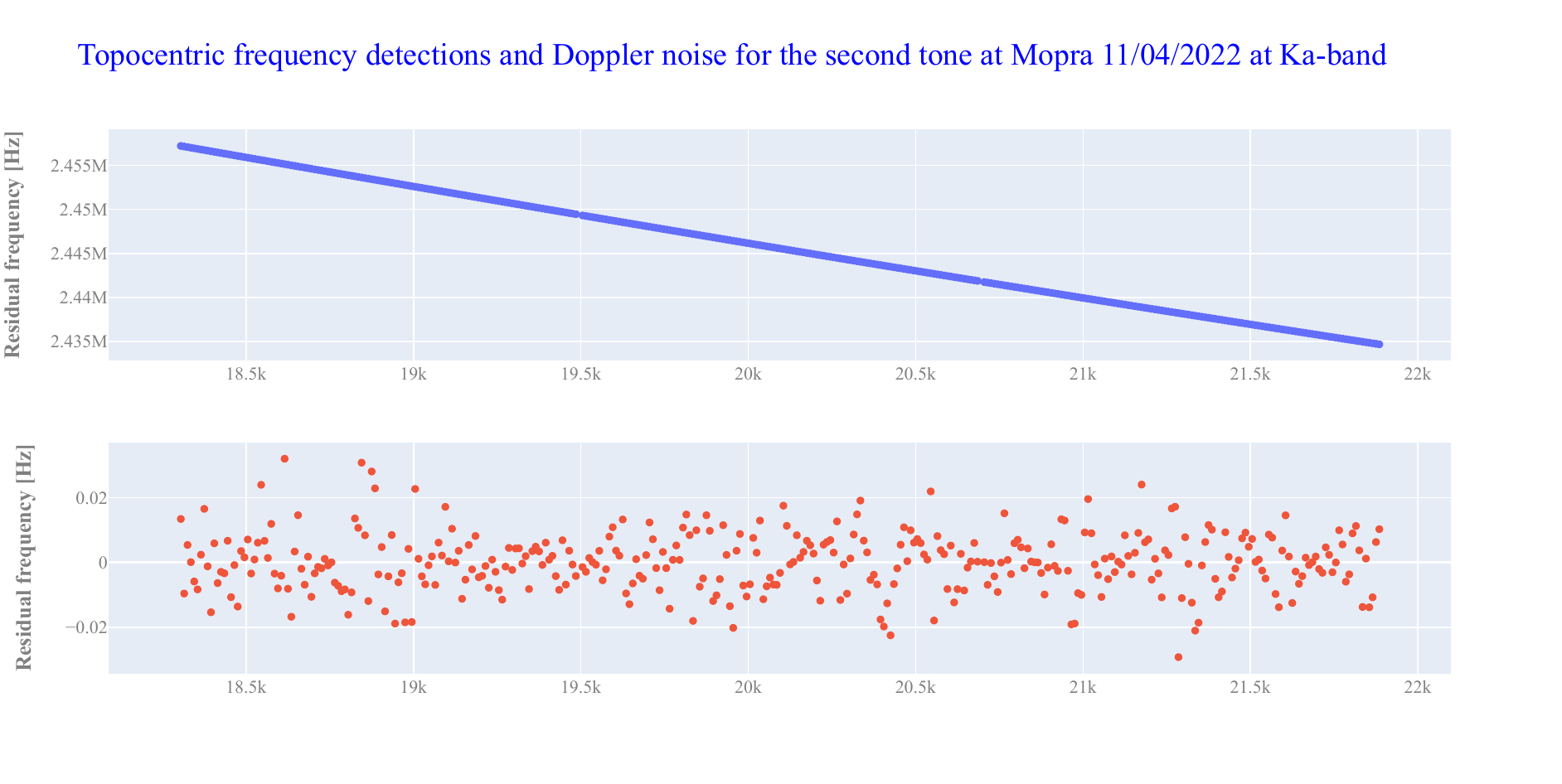}

    \caption{Topocentric frequency detections and Doppler noise at K$_{a}$-band as detected at Mopra on April 11, 2022, three 20-minute-long scans that were alternated with reference sources.The horizontal axis represents time, measured in seconds since the beginning of the day on April 11, 2022. This axis delineates the duration of observations conducted at the Mopra facility. Each point along this axis corresponds to a specific moment in time during the observations. The vertical axis depicts the detected frequency in kHz at the K$_{a}$-band. This frequency axis illustrates the variations in the detected signal over time due to the Doppler effect and noise. The Doppler shift, totaling 22 kHz over the 3600-second observation period, is represented on this axis.}
\label{fig:Mopra}
\end{figure*}

 To create and distribute schedules to VLBI telescopes and correlator, and organize an observation campaign for the discovered opportunity, the {custom-developed} planning tool's output is loaded into pySCHED. Many VLBI scheduling tools, such as pySCHED, have difficulty incorporating satellites as observation targets in the necessary control files, since these tools have been used for spacecraft observations only on relatively few occasions. Our setup allows operational VLBI data acquisition and processing systems to be used for direct satellite signal observations, similar to those used for operational radio source observations. pySCHED will provide elevation--time plots for each of the observing radio telescopes, among other things. This is important for scheduling the observational window and selecting the participating telescopes. 

\subsection{Tests with BepiColombo spacecraft }
\label{bepi}


PRIDE can, in principle, operate with radio signals in all frequency bands used in spacecraft communications and tracking. However, until recently, the technique had only demonstrated results at S- and X-bands. We have used the tools and methodology from Sections~\ref{develop} and \ref{result1} to plan, schedule and execute the BepiColombo observations at the K$_{a}$-band as a test case. The key objective of these experiments is to serve as a dry run and preparatory activity for the more extensive PRIDE--JUICE operations. As JUICE's and BepiColombo's radio communication systems and radio science instruments are very similar, BepiColombo is the most suitable choice for testing and preparations. 

The primary objective of the experiment was to detect the BepiColombo radio signal simultaneously at X- and K$_{a}$-bands.
Over 20 epochs have been observed between 2021 and 2022. In the specified epoch we consider here, we conducted a session in parallel using the radio telescopes from University of Tasmania (UTAS) Ceduna (Cd) in South Australia, Katherine (Ke) in North Territory, and the Commonwealth Scientific and Industrial Research Organisation (CSIRO) telescope at Mopra (Mp) in New South Wales. 

Figure~\ref{fig:Bepicolombo_PRIDE} shows the BepiColombo-PRIDE experiment settings, the Molleweide projection, and the zoomed-in finding chart. Figure~\ref{fig:ScVisibility_20220411} shows the SCHED elevation time plot for the telescopes used in this campaign, as an illustration of the output of the tools we have developed. The Cd and Ke tracked the downlink signal of the spacecraft at X-band, while Mopra tracked the radio signal at K$_{a}$-band. The spacecraft was operating in two-way mode with the ESTRACK ground station at New Norcia (NNO) in Western Australia. The uplink frequency was locked with an X/K$_{a}$ transponder on board (locked at 7.1 GHz).  We generated the schedule files for all three antennas, and we recorded data in VLBI Data Interchange Format (VDIF) for Ke and Cd, and LBA data format for Mp. The topocentric frequency detections at Mopra are shown in Fig.~\ref{fig:Mopra}. As shown in Fig.~\ref{fig:ScVisibility_20220411}, the observation was at a lower elevation; therefore, we only obtained 3 Mopra scans. The session was extended for three scans of 20 minutes. The Doppler noise or stochastic noise for the session was in the range of 10 mHz (as in Fig.~\ref{fig:Mopra}). 

{This section shows the results of an observation campaign that was performed using our planning and scheduling software for the specific case of BepiColombo. These specially designed modules aid in the meticulous scheduling to make sure the spacecraft is in the observable region of the telescopes during the planned observation time. Furthermore, the software makes it simple and feasible to modify the observation plan in (near) real-time when unforeseen changes in spacecraft operations take place. These modules enable easy collaboration between mission operators, ground-based observatories, and VLBI correlators to ensure successful and effective observations. For our science operations in the Jovian system, the VLBI data quality will be of the utmost importance to maximize PRIDE's impact on the science data products. Our first opportunity for an end-to-end test of our software and methodology, up to and including the processing of the VLBI data for use in improved ephemerides, will be the JUICE Venus flyby in 2025. In Section \ref{tudat}, we will present the results of a numerical simulation to quantify how VLBI data quality -- and, therefore, proper planning and scheduling -- will propagate into science data product quality.}

\subsection{Observing potential VLBI calibrator sources in preparation for the JUICE Venus flyby }
\label{venus-flyby}

JUICE will perform a Venus flyby (experiment PRIDE-EXP2; see Table~\ref{tab:experiments} and Section~\ref{cruise}) that will take place on 2025 Aug 31 (Table~\ref{tab:cruisephase}). Our plan to perform PRIDE observations (when JUICE is transmitting a radio signal before, during or after this flyby) has a number of reasons. Apart from offering a unique opportunity for a full dress rehearsal of PRIDE operations to be conducted later in the Jovian system, PRIDE-EXP2 has a potential scientific value in its own right, by promising an accurate independent validation of the ephemeris solution for Venus (see Section~\ref{tudat} for more details).  To improve the radio background near the Venus flyby, a preparatory VLBI experiment has been designed and conducted with the EVN (project EP129, PI: K. Perger). 

\begin{figure}
\centering
\includegraphics[scale=0.53]{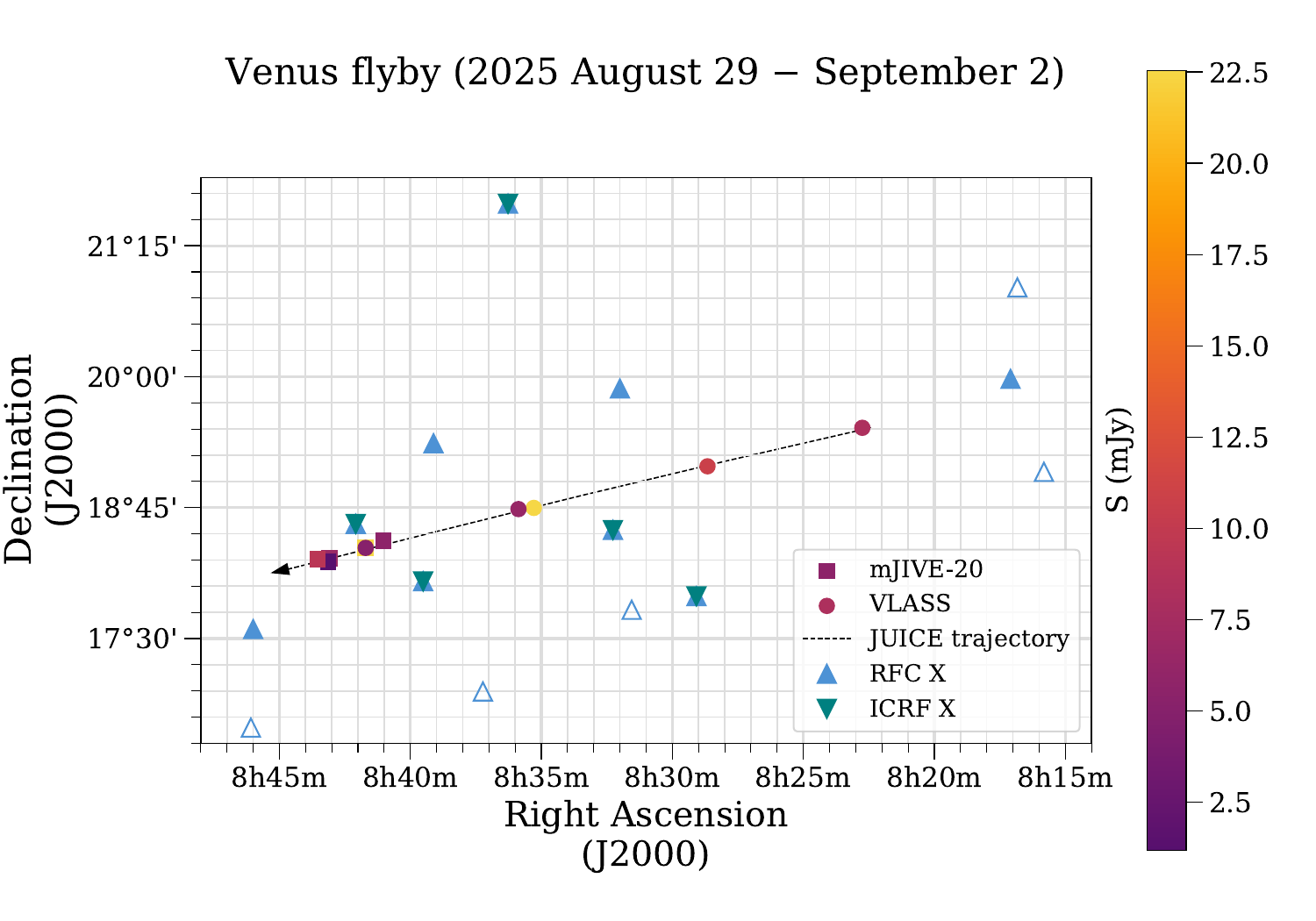}
\caption{Radio sources from the mJIVE-20 and VLASS catalogues within $4^{\prime}$ of the trajectory of JUICE (indicated by a dashed line) during 5 days around the Venus flyby. Their total flux densities at the L band (1.4~GHz) are coded with colours. Known VLBI calibrator sources from the ICRF and/or RFC within $\sim2^{\circ}$ are also shown. Sources marked by filled symbols are potential phase-reference calibrators in in-beam (circles and squares) or nodding styles (triangles). The list of their coordinates is given in Table~\ref{tab:calibrators}.}
\label{fig:venus-targets}
\end{figure}

While the quality of the PRIDE science products depends on VLBI observations of nearby phase calibrator sources \cite{rioja2020precise} with positions accurately linked to ICRF \cite{charlot2020third}, there are only 4 known ICRF sources found within $2^{\circ}$ of the spacecraft trajectory during the 5-day period centred on the date of the Venus flyby (Fig.~\ref{fig:venus-targets}). There are further 9 known VLBI-detected sources at X band in this celestial area found in the RFC \cite{PetrovL2022}. To use some of the latter as reference objects for JUICE VLBI measurements, their coordinates have to be linked to ICRF by means of relative astrometric observations (nodding-style phase referencing) to the neighbouring ICRF sources. For this purpose, we selected 3 RFC sources (J0832+1953, J0839+1921, and J0846+1735; see Table~\ref{tab:calibrators})) within $1^{\circ}$ of the JUICE trajectory. These sources are linked to two different neighbouring ICRF sources and supplement well the existing ICRF sources in the area, serving as nodding-style phase-reference sources for PRIDE-EXP2. Another RFC source, J0817+1958, would be an important potential phase-reference object at the beginning of the period where the sky chart in Fig.~\ref{fig:venus-targets} is sparsely populated with known VLBI sources, but unfortunately no ICRF sources are found within $2^{\circ}$ from its position. Therefore, the current radio background of JUICE near the Venus flyby could limit the quality of the VLBI data that could be acquired. Our preparatory experiment will mitigate this issue.

\begin{table}[h]
\caption{Known and potential calibrator sources for PRIDE-EXP2 to be conducted at around the JUICE Venus flyby in 2025. The five ICRF sources (top) and four additional RFC sources (middle) have already been detected with VLBI at X band, while the suitability of nine potential in-beam calibrators (bottom) needs to be verified in a dedicated preparatory VLBI experiment. A sky chart with the source positions is shown in Fig.~\ref{fig:venus-targets}. }\label{tab:calibrators}
\begin{tabular*}{\textwidth}{@{\extracolsep\fill}lllcc}
\toprule
\textbf{Source Name} & \textbf{Right Ascension} & \textbf{Declination} & \textbf{Estimated Correlated} & \textbf{Reference} \\
            & \textbf{(h min s)}         & \textbf{($\circ$ $\prime$ $\prime\prime$)} & \textbf{Flux Density (mJy)} & \\
\midrule
J0829+1754   & 08 29 04.828611  & +17 54 15.86399  & 30 & ICRF \cite{charlot2020third} \\
J0832+1832   & 08 32 16.040283  & +18 32 12.13313  & 50 & ICRF \cite{charlot2020third} \\
J0836+2139   & 08 36 16.216892  & +21 39 03.58055  & 100 & ICRF \cite{charlot2020third} \\
J0839+1802   & 08 39 30.721364  & +18 02 47.14274  & 120 & ICRF \cite{charlot2020third} \\
J0842+1835   & 08 42 05.094174  & +18 35 40.99051  & 200 & ICRF \cite{charlot2020third} \\
\midrule
J0817+1958$^{*}$   & 08 17 05.49328  & +19 58 42.8985  & 30 & RFC \cite{PetrovL2022} \\
J0832+1953$^{*}$   & 08 32 00.15804  & +19 53 12.1757  & 40 & RFC \cite{PetrovL2022} \\
J0839+1921$^{*}$   & 08 39 06.94389  & +19 21 48.6850  & 30 & RFC \cite{PetrovL2022} \\
J0846+1735$^{*}$   & 08 46 00.03366  & +17 35 25.1140  & 20 & RFC \cite{PetrovL2022} \\
\midrule
J0822+1930$^{*}$   & 08 22 44.877  & +19 30 43.40   & 2$^{\dagger}$  & VLASS \cite{gordon2021vlass} \\
J0828+1908$^{*}$   & 08 28 39.432   & +19 08 40.64   & 2$^{\dagger}$  & VLASS \cite{gordon2021vlass} \\
J0835+1844A$^{*}$  & 08 35 17.235   & +18 44 50.96   & 6$^{\dagger}$  & VLASS \cite{gordon2021vlass} \\
J0835+1844B$^{*}$ & 08 35 52.393   & +18 44 10.02   & 2$^{\dagger}$  & VLASS \cite{gordon2021vlass} \\
J0841+1821$^{*}$   & 08 41 42.3106  & +18 21 56.774  & 2$^{\dagger}$  & mJIVE \cite{deller2014mjive}, \\
 & & & & VLASS \cite{gordon2021vlass} \\
J0841+1826$^{*}$   & 08 41 01.6482  & +18 26 02.421  & 1$^{\dagger}$  & mJIVE \cite{deller2014mjive} \\
J0843+1816$^{*}$   & 08 43 08.5839  & +18 16 05.078  & 1$^{\dagger}$  & mJIVE \cite{deller2014mjive} \\
J0843+1814$^{*}$   & 08 43 08.5839  & +18 14 00.373  & 1$^{\dagger}$  & mJIVE \cite{deller2014mjive} \\
J0843+1815$^{*}$   & 08 43 32.5761  & +18 15 31.062  & 1$^{\dagger}$  & mJIVE \cite{deller2014mjive} \\
\botrule
\end{tabular*}
Notes: $^{*}$ -- accurate position to be linked to ICRF; $^{\dagger}$ -- VLBI detectability to be checked in the EVN experiment EP129 
\end{table}

The half-power width of the primary beam of a 32-m diameter radio telescope is $\sim 4^\prime$ at X band. Therefore, identifying suitable calibrator sources within $4^\prime$ of JUICE trajectory would allow in-beam phase-referencing observations before and/or after the Venus flyby with a network of $\sim32$-m or smaller antennas. To search for candidate objects, we checked the mJy Imaging VLBA Exploration at 20 cm (mJIVE-20) \cite{deller2014mjive} list containing VLBA-detected sources at 1.4~GHz, and the second-epoch Very Large Array Sky Survey (VLASS) \cite{gordon2021vlass} catalogue at 3~GHz. The latter sources are only detections with a short-baseline connected-element interferometer, the Karl G. Jansky Very Large Array (VLA), having yet unknown milliarcsecond-scale structure and VLBI detectablity. There are a total of $12$ radio sources, with flux densities up to about $24$~mJy, along the JUICE trajectory in mJIVE-20 and/or VLASS. Out of them, $3$ sources were found to have extended lobe-dominated morphology on arcsecond scale and thus dropped from the sample. The remaining $9$ sources (Fig.~\ref{fig:venus-targets}, Table~\ref{tab:calibrators}) can be considered as candidate in-beam phase reference calibrators. Their detectability with VLBI and suitability as calibrators for JUICE have to be checked by means of prior X-band imaging observations. For the candidate in-beam reference source J0822+1930, we planned phase-referencing observations to the closest ICRF source (J0829+1754, $2.26^{\circ}$) as well as the RFC source J0817+1958 mentioned before as potential phase-reference source in the period preceding the Venus flyby. This way, upon successful detection of J0822+1930, the position of J0817+1958 could indirectly be linked to ICRF. The $10$-h preparatory EVN experiment was conducted on 2024 Feb 23--24. At the time of writing, the data are being prepared for correlation. After correlation, calibration and data analysis, the results will become available for planning PRIDE-EXP2. In particular, this experiment will improve the quality, diversity and density of phase-reference sources, thereby allowing more accurate VLBI observations of JUICE to be acquired close to the Venus flyby. The potential scientific implications of this are discussed in the next section.

\subsection{Outlook to JUICE Venus flyby }
\label{tudat}

Improving planetary and satellite ephemerides is a key objective of PRIDE for the JUICE mission. As mentioned in Section~\ref{develop}, the Venus flyby performed by the JUICE spacecraft during its cruise phase will represent an excellent opportunity to test the PRIDE technique, the data analysis procedures to provide input data points to ephemerides, and quantify the improvement that PRIDE observables can bring to ephemerides solutions (PRIDE-EXP2, Table~\ref{tab:cruisephase}). 
We will be able to provide a very accurate normal point (local state solution) for Venus at the flyby epoch, which will help us verify our methodology, and offers an additional validation opportunity for Venus' ephemeris. In this section, we simulate this Venus flyby experiment and analyse the influence of accuracy levels for PRIDE VLBI measurements on Venus' state solution. This provides a direct quantification of the impact of scheduling and planning on science return. Since the science return (in terms of ephemeris improvement) from PRIDE will be made primarily from data acquired at the flybys \citep{fayolle2024}, the Venus flybys are of prime interest. 

To quantify the accuracy of the local state solution achievable for Venus, we used the Tudat\footnote{https://docs.tudat.space/} software \citep{dirkx2019propagation} to perform a covariance analysis. We simulate PRIDE Doppler data \citep[every 60~s, with a noise level of 35~$\mu$m\,s$^{-1}$;][]{bocanegra2018planetary}, as well as independent VLBI measurements every 20 minutes. We assumed a tracking arc of 2~h centered at the closest approach. We considered different VLBI noise levels, adopting identical PRIDE VLBI error budgets\footnote{Poor VLBI case: 0.6 and 0.9 nrad noise level in right ascension and declination, respectively. Good VLBI case: 0.2 and 0.3 nrad in right ascension and declination. In-beam VLBI case: 0.1 nrad in both right ascension and declination.} as in \citep{fayolle2024}, to test the impact of various data quality scenarios on the state estimation \citep[see similar analyses in][]{fayolleEtAl2022,fayolle2023}. The detailed descriptions of VLBI simulations for the Jovian system are provided in \citep{fayolle2024}. Different phase calibrators identified in Table \ref{tab:calibrators} were investigated, their ICRF position uncertainty being included as a systematic error (i.e., bias) to the VLBI observables \citep{fayolle2024}, and added as consider parameters in our analysis (see below). More precisely, we selected the four RFC calibrators identified within 2 degrees of JUICE, and two ICRF ones, picked such that we sample a representative range of ICRF position errors (see Table \ref{tab:VenusStateSolutions}). Potential in-beam calibrators listed in Table \ref{tab:calibrators} were not selected since their suitability as phase referencing sources is yet to be confirmed (see Section~\ref{venus-flyby} about the selection of potentially suitable in-beam calibrators). Comparing the results of the different cases will provide information on the added value of expanding more effort to improve observation planning.

From the simulated PRIDE observables, we estimated both JUICE’s and Venus’ states at the flyby epoch. No \textit{a priori} constraint was applied to the state of the JUICE spacecraft, while \textit{a priori} uncertainties for Venus’ state were set to three times the differences between the INPOP21 \citep{fienga2021} and DE440 \citep{park2021jpl} ephemerides solutions, used as a conservative estimate of Venus' ephemeris error. This led to \textit{a priori} constraints of 10 m (radial), 480 m (tangential) and 630 m (normal), the radial position being very well constrained by the Magellan and Venus Express data.

\begin{table}[h]
\caption{Formal uncertainties in Venus' position at the time of JUICE flyby using different phase calibrators and VLBI noise levels. The VLBI biases are defined by the calibrator' position uncertainties, reported in the second and third columns. The resulting Venus' state solutions are provided in last three columns, in RTN (Radial; Tangential; Normal) coordinates.}\label{tab:VenusStateSolutions}
\begin{tabular*}{\textwidth}{@{\extracolsep\fill}lcccccc}
\toprule%

Phase calibrator & $\sigma(\alpha)$ [mas] & $\sigma(\delta)$ [mas] & VLBI noise & $\sigma(R)$ [m] & $\sigma(T)$ [m] & $\sigma(N)$ [m] \\
\midrule
\multirow{3}{*}{J0817+1958} & \multirow{3}{*}{0.19} & \multirow{3}{*}{0.23} & poor & 10 & 260 & 250 \\
& & & good & 10 & 260 & 240 \\
& & & in-beam & 10 & 260 & 240 \\
\midrule
\multirow{3}{*}{J0832+1953} & \multirow{3}{*}{0.59} & \multirow{3}{*}{0.92} & poor & 10 & 840 & 960 \\
& & & good & 10 & 840 & 960 \\
& & & in-beam & 10 & 840 & 960 \\
\midrule
\multirow{3}{*}{J0839+1921} & \multirow{3}{*}{0.16} & \multirow{3}{*}{0.25} & poor & 10 & 230 & 270 \\
& & & good & 10 & 230 & 260 \\
& & & in-beam & 10 & 230 & 260 \\
\midrule
\multirow{3}{*}{J0829+1754} & \multirow{3}{*}{0.14} & \multirow{3}{*}{0.19} & poor & 10 & 200 & 210 \\
& & & good & 10 & 200 & 200 \\
& & & in-beam & 10 & 200 & 200 \\
\midrule
\multirow{3}{*}{J0842+1835} & \multirow{3}{*}{0.11} & \multirow{3}{*}{0.11} & poor & 10 & 160 & 140 \\
& & & good & 10 & 150 & 120 \\
& & & in-beam & 10 & 150 & 110 \\
\midrule
no VLBI & & & & 10 & 480 & 630 \\
\botrule
\end{tabular*}
\end{table}

In addition to JUICE’s and Venus’ states, we added a number of considered parameters to ensure that their current uncertainties are accounted for in our estimations \citep{montenbruck2002}. Venus’ gravitational parameter, spherical harmonic gravity field coefficients up to degree and order 10 rotation rate, and pole orientation were included, and their uncertainties were based on the formal uncertainties derived from Magellan data analysis \citep{konopliv1999} and ground-based observations \citep{margot2021}. A factor of 3 was, however, systematically applied to the published formal errors, as suggested in \citep{konopliv1999}. Empirical accelerations acting on the spacecraft, estimated every hour during tracking, with a consider uncertainty of $1.0\times10^{-8}$~m\,s$^{-2}$. As a sensitivity analysis, we have verified that our results for the Venus state change minimally when we use an \textit{a priori} value that is 10 times higher. Finally, VLBI biases were added with a consider uncertainty equal to the calibrator position uncertainty.

Table~\ref{tab:VenusStateSolutions} displays the formal errors in Venus’ position obtained for various VLBI random noise levels and calibrator position uncertainties. It must be noted that, as expected, Doppler measurements do not noticeably contribute to Venus’ state solution, which remains constrained to their \textit{a priori} values when no VLBI is included. Adding VLBI, however, significantly reduces the \textit{a priori} uncertainties in the tangential and normal directions. Only Venus’ radial position cannot be improved beyond its \textit{a priori} uncertainties (irrespective of the error budget). It is interesting to note that the influence of the PRIDE VLBI random noise level is rather limited. The choice of phase calibrator, on the other hand, is critical: the accuracy with which its ICRF position is known, by directly affecting the systematic error in the VLBI measurement of JUICE's angular position, drives the quality of the resulting state solution for Venus. These results provide a very strong motivation for the thorough analysis of available calibrators, including the possibility to use multiple calibrators or performing a dedicated campaign to extend/enhance the available calibrator sources. In particular, reducing the systematic bias in VLBI measurement induced by the error in the calibrator's position is essential. The tools developed for PRIDE and presented here allow analyses of the feasibility and scientific return of such activities to be performed with minimal effort.


\section{Summary and conclusion}
\label{summary}

PRIDE, as one of eleven JUICE experiments, aims to enhance the mission's science return by applying VLBI and Doppler tracking techniques.
In 2031, JUICE spacecraft will arrive in the Jovian system and conduct extensive observations of Ganymede and flyby observations of Europa and Callisto. The two primary goals of PRIDE are to improve the ephemerides for the Jovian system and study the Jovian atmosphere and the ionosphere of the Galilean moons using radio occultation observations. 

In this work, we have reviewed the PRIDE operations
and engineering interactions with the JUICE spacecraft.
Furthermore, a step-by-step procedure was presented
for planning and scheduling
PRIDE observation campaigns in a manner that automates many of the steps in the process. 
By selecting the celestial regions of interest for JUICE's near-field VLBI (PRIDE) during various mission stages, notably during the cruise period, we showcase the significance of scheduling and its impact on the experiment's science return. For JUICE, a finding chart for extragalactic radio sources from the Radio Fundamental Catalogue was created. 

The ESA's BepiColombo spacecraft was observed 
as a training target for prospective optional K$_{a}$-band PRIDE operations. The initial results on PRIDE K$_{a}$-band Doppler shift detections are demonstrated. This example demonstrates the flexibility of the scheduling approach and PRIDE's readiness for operations on any planetary mission, which proved to enhance the process and boost the scientific return.

Additionally, we have presented a numerical simulation of the PRIDE contribution to potential science return from Venus flyby, in which we took into account different tracking durations and noise levels for VLBI observations. This allowed us to test various acquisition and error budget scenarios while also, and perhaps more importantly, quantifying their impact on the final ephemeris solution. The simulations also demonstrated that by lowering PRIDE VLBI noise (for instance, using K$_{a}$-band or in-beam tracking), one would further improve the state solution for the flyby body. Additionally, choosing a phase calibrator is crucial; its ICRF position accuracy directly influences the error in measuring JUICE's position through VLBI, thus impacting the accuracy of determining the Venus state solution. These results strongly recommend an examination of calibrators, which may involve using multiple ones or organizing specific campaigns to expand/improve the calibrator options. It is essential to reduce biases in VLBI measurements caused by errors in calibrator positions.
This indicates the need for effective planning and scheduling of PRIDE--JUICE observations during the Galilean moon flyby tour to ensure maximum VLBI data quality and, as a result, provide the best possible scientific contribution of the PRIDE experiment in improving our understanding of icy satellite origins, interior and evolution.

\section*{Acknowledgments}
The EVN is a joint facility of independent European, African, Asian, and North American radio astronomy institutes. The involvement of the University of Tasmania in the observations presented here was enabled by Geoscience Australia and the Australian Government via the National Collaborative Research Infrastructure Strategy (NCRIS). J.F., S.F., and K.P. acknowledge the ESA PRODEX support (project PEA 4000136207). We express our gratitude to Olivier Witasse, Claire Vallat, Nicolas Altobelli, and the ESA team (MOC and SOC) for the cooperation provided for supporting PRIDE observations of JUICE and BepiColombo missions. We acknowledge Chris Phillips (CASS) for supporting and conducting the observations at Mopra. We appreciate the anonymous referees for their valuable comments and corrections.

\end{document}